\documentclass[acmlarge]{acmart}
\AtBeginDocument{%
  }

\usepackage[nolist]{acronym}
\usepackage{multirow}
\usepackage{soul}

\definecolor{change_color}{rgb}{0,0,1}

 \setcopyright{acmlicensed}
 \copyrightyear{2026}
 \acmYear{0}
 \acmDOI{XXXXXXX.XXXXXXX}

\acmJournal{IMWUT}
\acmVolume{0}
\acmNumber{0}
\acmArticle{0}
\acmMonth{0}

\begin{acronym}
    \acro{AUDIT}{alcohol use disorders identification test}
    \acro{AUROC}{area under the receiver operating characteristic curve}
    \acro{AUPRC}{area under the precision-recall curve}
    \acro{BAC}{blood alcohol concentration}
    \acro{CNN}{convolutional neural network}
    \acro{DMC}{driver monitoring camera}
    \acro{ECG}{electrocardiogram}
    \acro{HR}{heart rate}
    \acro{HRV}{heart rate variability}
    \acro{IBI}{interbeat interval}
    \acro{LOSO}{leave-one-subject-out}
    \acro{ML}{machine learning}
    \acro{PEth}{phosphatidylethanol}
    \acro{PPG}{photoplethysmography}
    \acro{ROC}{receiver operating characteristic}
    \acro{WHO}{World Health Organization}
\end{acronym}

\begin{document}

\title{Detecting Drunk Driving Using Off-the-Shelf Smartwatches}

\author{Robin Deuber}
\email{rdeuber@ethz.ch}
\orcid{0009-0007-6608-7208}
\affiliation{%
  \institution{ETH Z{\"u}rich}
  \city{Z{\"u}rich}
  \state{Z{\"u}rich}
  \country{Switzerland}
}

\author{Lanlan Yang}
\email{lanlan.yang@uzh.ch} 
\orcid{0009-0007-1582-1492}
\affiliation{%
  \institution{University of Z{\"u}rich}
  \city{Z{\"u}rich}
  \state{Z{\"u}rich}
  \country{Switzerland}
}

\author{Michal Bechny}
\email{mbechny@ethz.ch}
\orcid{0009-0006-0845-1844}
\affiliation{%
  \institution{ETH Z{\"u}rich}
  \city{Z{\"u}rich}
  \state{Z{\"u}rich}
  \country{Switzerland}
}

\author{Christoph Heck}
\email{hchristo@ethz.ch}
\orcid{0009-0008-6102-6040}
\affiliation{%
  \institution{ETH Z{\"u}rich}
  \city{Z{\"u}rich}
  \state{Z{\"u}rich}
  \country{Switzerland}
}

\author{Matthias Pf\"affli}
\email{matthias.pfaeffli@irm.unibe.ch}
\orcid{0000-0003-2712-8672}
\affiliation{%
  \institution{University of Bern}
  \city{Bern}
  \country{Switzerland}}

\author{Matthias Bantle}
\email{matthias.bantle@irm.unibe.ch}
\orcid{0000-0001-9210-2802}
\affiliation{%
  \institution{University of Bern}
  \city{Bern}
  \country{Switzerland}}

\author{Florian von Wangenheim}
\email{fwangenheim@ethz.ch}
\orcid{0000-0003-3964-2353}
\affiliation{%
  \institution{ETH Z\"urich}
  \city{Z\"urich}
  \country{Switzerland}}

\author{Elgar Fleisch}
\email{efleisch@ethz.ch}
\orcid{0000-0002-4842-1117}
\affiliation{%
  \institution{ETH Z\"urich}
  \city{Z\"urich}
  \country{Switzerland}}
\affiliation{%
  \institution{University of St. Gallen}
  \city{St. Gallen}
  \country{Switzerland}}

\author{Wolfgang Weinmann}
\email{wolfgang.weinmann@irm.unibe.ch}
\orcid{0000-0001-8659-1304}
\affiliation{%
  \institution{University of Bern}
  \city{Bern}
  \country{Switzerland}}

\author{Manuel G\"unther}
\email{siebenkopf@googlemail.com}
\orcid{0000-0003-1489-7448}
\affiliation{%
  \institution{University of Z{\"u}rich}
  \city{Z{\"u}rich}
  \state{Z{\"u}rich}
  \country{Switzerland}} 

\author{Felix Wortmann}
\authornote{Shared last authorship.}
\email{felix.wortmann@unisg.ch}
\orcid{0000-0001-5034-2023}
\affiliation{%
  \institution{University of St. Gallen}
  \city{St. Gallen}
  \country{Switzerland}}

\author{Varun Mishra}
\email{v.mishra@northeastern.edu}
\authornotemark[1]
\orcid{0000-0003-3891-5460}
\affiliation{%
  \institution{Northeastern University}
  \city{Boston}
  \state{Massachusetts}
  \country{USA}
}

\renewcommand{\shortauthors}{Deuber et al.}

\newcommand{\EARLY}[0]{\textsc{Early Warning}}
\newcommand{\ABOVE}[0]{\textsc{Above Limit}}

\newcommand{\random}[1]{\color{gray!60}#1}

\newcommand{\gdL}[1]{#1\,g/dL}
\newcommand{\mgL}[1]{#1\,mg/L} 
\newcommand{\kmh}[1]{#1\,km/h}

\newcommand{\meanstd}[2]{#1\,$\pm$\,#2}

\begin{abstract}

  Alcohol-impaired driving remains a major yet preventable cause of road traffic injury and death, with many drivers underestimating their level of intoxication. Compared to in-vehicle systems, mobile drunk-driving detection using consumer smartwatches offers a scalable way to trigger preventive interventions and increase awareness without additional in-vehicle hardware. We introduce a system that leverages wrist accelerometer data and heart rate variability-derived physiological signals to detect alcohol-related driving impairment. We collected data in a randomized, controlled three-arm test-track study ($n=54$) and trained both logistic regression models with window-aggregated features and a two-tower 1D convolutional neural network (CNN), to detect alcohol-impaired driving. The CNN achieved a participant-averaged area under the receiver operating characteristic (AUROC) of 0.88 for detecting any alcohol intoxication and 0.86 for detecting driving above the WHO-recommended limit of 0.05 g/dL. To the best of our knowledge, this is the first work to (1) demonstrate drunk-driving detection using consumer smartwatches, (2) develop and evaluate such a system in a real vehicle on a closed test track, and (3) rigorously assess generalization to unseen participants. Together, these findings highlight the potential of wearable-based sensing to support scalable, measurement-driven prevention of alcohol-related traffic harm.
\end{abstract}




\maketitle

\section{Introduction}
\label{secC:introduction}

\begin{figure}[htbp]
  \centering
  \includegraphics[width=\linewidth]{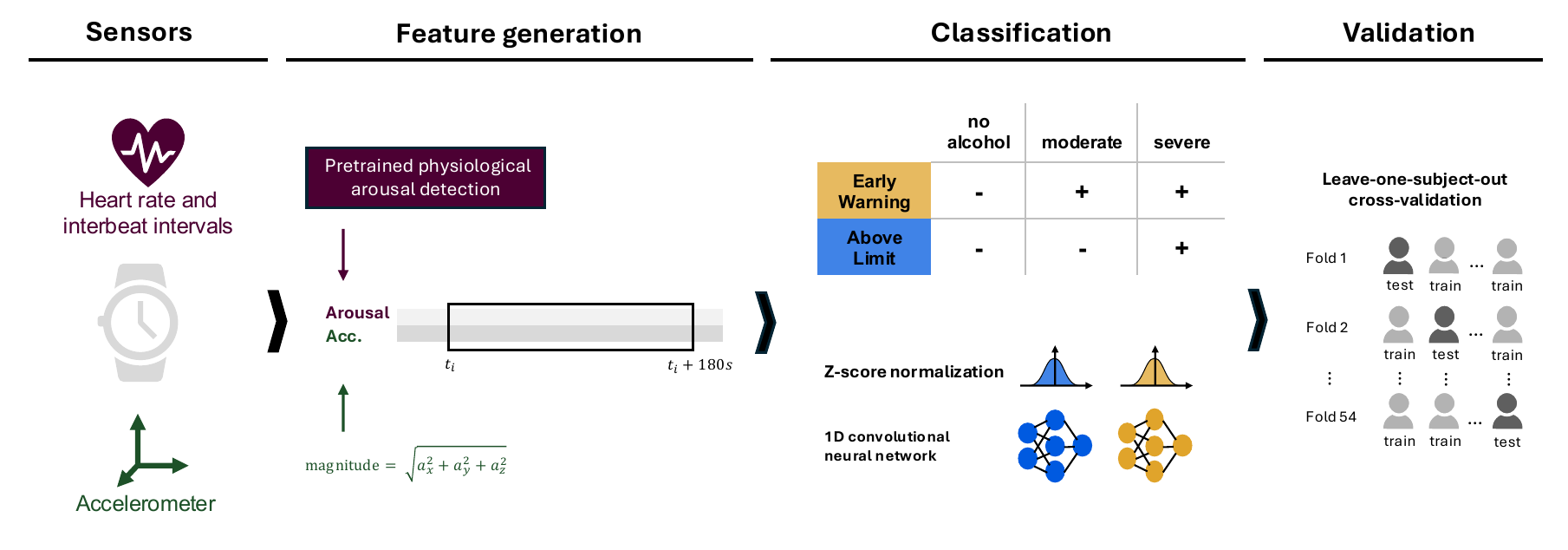}
  \caption{Overview of the 1D CNN pipeline designed to perform two classification tasks (\EARLY{} and \ABOVE{}).}
  \Description{} 
  ~\label{figC:pipeline}
\end{figure}

According to a 2024 report by the \ac{WHO}, alcohol consumption caused approximately 2.6 million deaths worldwide in 2019 and represents a major modifiable risk factor contributing substantially to the global burden of disease and injury, with impacts extending across health, safety, and social domains~\cite{who2024alcohol}.
In the context of road safety, alcohol-related road crashes caused approximately 298{,}000 deaths worldwide in 2019, of which 156{,}000 fatalities involved individuals who had not consumed alcohol themselves, highlighting that alcohol consumption poses substantial risks not only to one's own health but also to the safety of others~\cite{who2024alcohol}.

Drunk driving generally refers to operating a motor vehicle while alcohol consumption has impaired the cognitive, perceptual, or motor abilities required for safe driving. Empirical evidence shows that alcohol-related crash risk increases nonlinearly with intoxication level: a recent meta-analysis pooling over 100 effect estimates found an exponential relationship between \ac{BAC} and the risk of being killed or seriously injured, with odds ratios reaching up to approximately 240 for \ac{BAC} levels between 0.12 and 0.20~g/dL~\cite{hoye2023alcohol}. 
Despite this continuous risk increase, legal \ac{BAC} thresholds vary substantially across regions, including zero tolerance policies as well as limits such as 0.05 and 0.08~g/dL~\cite{DrinkDrivingLimitsWW}. 
In practice, drunk driving most frequently occurs in everyday situations following social drinking, such as evenings out or celebratory events, where individuals may underestimate their level of impairment or overestimate their ability to compensate. While alcohol-impaired driving occurs across all demographic groups, it is disproportionately prevalent among young adults, particularly those aged 21--24, and among men, consistent with broader patterns of social drinking and risk-taking behavior shaped by situational constraints and social norms~\cite{NHTSATrafficSafetyFacts,CDCImpairedDriving}.

Alcohol consumption induces pronounced physiological changes that affect how the body regulates stress, attention, and motor control. 
In particular, acute alcohol intake disrupts autonomic nervous system regulation, leading to reduced \ac{HRV}~\cite{weise1986acute, reed1999effect, romanowicz2011changes, brunner2021impact, koskinen1994acute}. 
Beyond autonomic effects, alcohol also impairs sensorimotor integration, disrupting gross motor control and movement smoothness~\cite{modig2012blood, dalrymple2003effects}. 
These physiological and motor disturbances compromise cognitive, visual, and fine motor functions essential for safe driving, including sustained attention, decision-making, hazard perception, and precise motor coordination~\cite{moskowitz2000review, martin2013review}.
Consequently, intoxicated drivers respond more slowly to unexpected events, exhibit degraded lane keeping (e.g., increased lateral variability), show increased steering variability, and display impaired visual-motor coordination~\cite{calhoun2004alcohol, irwin2017effects, garrisson2022effects}. 
Together, this evidence demonstrates that alcohol-related physiological dysregulation manifests as measurable deficits in visual control, movement smoothness, and driving performance that substantially increase crash risk~\cite{hoye2023alcohol, CDCImpairedDriving, NHTSATrafficSafetyFacts, EC2024AlcoholDrugs}.

Beyond these physiological and behavioral impairments, alcohol intoxication can compromise self-awareness, and many drinkers struggle to estimate their level of intoxication and fitness to drive accurately. Experimental and field studies similarly show substantial miscalibration between subjective and objective intoxication, and \ac{BAC} underestimation has been linked to riskier driving behavior~\cite{kochling2021hazardous,laude2016drivers}. Acute alcohol tolerance can further reduce perceived danger and increase the willingness to drive \cite{amlung2014effects}. At the same time, impaired driving also persists due to a compliance gap: some individuals drive despite being aware of impairment or legal risk. The present work focuses on the former mechanism (miscalibration and reduced situational awareness) rather than on deliberate rule violations. 

Objective impairment detection could therefore support accident prevention in two complementary ways. First, timely feedback could increase awareness where individuals might otherwise decide to continue driving despite impairment. Second, modern vehicles increasingly incorporate driver-state-aware assistance systems that could account for the driver’s condition into their control logic, for example, by initiating emergency braking earlier~\cite{wang2015forward} and thereby increasing safety. Despite these opportunities, closing the awareness gap at scale requires sensing that is both objective and broadly accessible. Many existing approaches require in-vehicle or specialized sensors, or proprietary integration that is not available across the broader vehicle fleet (e.g. \cite{koch2023leveraging, Deuber2025}), which can limit coverage and raise practical deployment barriers.

These considerations motivate wearable-based drunk-driving detection as a more accessible, non-invasive, and potentially rapid solution. Wrist-worn wearables are widely adopted and continue to ship at scale~\cite{eurostatWearables}, and driving is a highly structured activity characterized by repetitive control actions that are observable at the wrist. Alcohol is known to affect autonomic regulation and motor behavior (e.g., reduced \ac{HRV}, elevated physiological arousal, and altered movement patterns) in ways that are directly measurable using consumer-grade wearable devices via \ac{HRV} features and accelerometry~\cite{koskinen1994acute, romanowicz2011changes, brunner2021impact}. In summary, the rationale for smartwatch-based detection is not that all intoxicated drivers would necessarily refrain from driving once notified, but rather that a meaningful subset of impaired-driving episodes arises from miscalibration of one’s own intoxication level and fitness to drive. In such cases, objective feedback delivered through a widely available wrist-worn device may help reduce this awareness gap and support preventive decision-making; while the same signal could also serve as an input to driver-state-aware vehicle safety systems.

Despite these factors and the potential to reduce the awareness gap among drivers, it remains unclear whether alcohol-impaired driving can be detected using off-the-shelf smartwatches in a real driving context. To address this gap, we introduce a smartwatch-based drunk-driving detection approach that leverages only physiology and accelerometer signals to identify alcohol-related impairment during real driving. In particular, we investigate to what extent alcohol-impaired driving can be detected in a real vehicle using off-the-shelf smartwatches, and assess the extent to which the resulting models generalize to unseen drivers.

In summary, this work makes three main contributions. First, we demonstrate that alcohol-impaired driving can be detected using off-the-shelf smartwatches that sense accelerometer and \ac{HRV} data, without requiring in-vehicle telemetry or driver-facing cameras. Second, we develop and evaluate such a system in a real vehicle on a closed test track, thereby moving wearable-based intoxication detection from unconstrained or gait-based settings to a structured driving context in which gait cues are unavailable and wrist movements are shaped by vehicle control. Third, we rigorously assess generalization to unseen drivers via \ac{LOSO} validation and provide a structured comparison of a logistic-regression baseline and a two-tower 1D \ac{CNN}, including analyses of (temporal context) window length, modality ablations, and per-driving-phase normalization (see Figure \ref{figC:pipeline}).



In the spirit of open science and reproducibility, we release the source code accompanying this paper at \url{https://anonymous.4open.science/r/Detecting-Drunk-Driving-Using-Off-the-Shelf-Smartwatches/}. In accordance with local ethics requirements, de-identified study data may be made available for non-commercial research upon reasonable request to the authors, subject to approval by the scientific study board and a data transfer agreement.

\section{Related Work}
\label{secC:related_work}

\subsection{Wearable-Based Detection of Alcohol Intoxication}
\label{secC:wearable_alcohol_intoxication_detection}

Alcohol intoxication can be assessed through a wide range of sensing modalities \cite{paprocki2022review}. These include approaches that estimate \ac{BAC} from reported consumption and physiological characteristics; breath analyzers; bodily fluid analysis (e.g., blood analysis via gas chromatography); transdermal sensing of ethanol or its metabolites; optical spectroscopy that detects ethanol signatures in tissue; and indirect inference based on physiological or behavioral signals (e.g., \ac{PPG}, body temperature, or driving behavior).

Focusing on wearable sensing, Davis-Martin et al.~\cite{davis2021alcohol} categorize biosensor-based intoxication detection into gait-based assessment (e.g., accelerometer and gyroscope) and transdermal alcohol concentration sensing. Although transdermal alcohol concentration sensors directly quantify ethanol diffused through the skin, they are not integrated into consumer-grade smartwatches. \NoHyper\citeauthor{brobbin2022accuracy}\endNoHyper\ identified 32 transdermal alcohol con\-cen\-tra\-tion-focused wearable studies but reported significant methodological variability \cite{brobbin2022accuracy}. 

Recent work by \NoHyper\citeauthor{FAIRBAIRN2025112519}\endNoHyper\ achieved up to 0.94 \ac{AUROC} in intoxication detection using a wrist-worn transdermal alcohol biosensor evaluated across both controlled laboratory sessions and real-world field use~\cite{FAIRBAIRN2025112519}. In contrast, \NoHyper\citeauthor{panneer2016wearable}\endNoHyper\ demonstrated a wearable sweat-based biochemical sensor for detecting an ethanol metabolite~\cite{panneer2016wearable}. Because metabolites can remain detectable substantially longer than contemporaneous intoxication, sweat-based systems are primarily suited to monitoring alcohol consumption over longer time horizons rather than real-time impairment at the moment of driving~\cite{panneer2016wearable}.

Several studies explored physiological sensing for intoxication detection outside of the driving context. \NoHyper\citeauthor{chen2018non}\endNoHyper\ proposed a non-invasive sobriety-test system based on finger \ac{PPG}, and reported an accuracy of up to 85\% for detecting alcohol-related changes in \ac{PPG} signals~\cite{chen2018non}. \NoHyper\citeauthor{wang2018svm}\endNoHyper\ developed an intoxication-identification system using \ac{ECG} and \ac{PPG} sensors, and trained SVM classifiers, reporting an average identification performance of 95\% in accuracy/F1-score; notably, they found \ac{PPG}-based features to perform comparably to \ac{ECG} while being more convenient to acquire~\cite{wang2018svm}. Fingertip \ac{PPG} has also been explored in small-scale prototypes using reflected red/infrared optical sensing and breath-alcohol reference measurements, achieving up to 87.5\% classification accuracy between alcohol-consumed vs. non-consumed conditions~\cite{sanguansri2021development}. Beyond optical sensing, \NoHyper\citeauthor{czaplik2019evaluation}\endNoHyper\ evaluated bioimpedance spectroscopy and impedance cardiography in a controlled drinking trial (ethanol vs.\ water control), using repeated breath-alcohol measurements and showing that bioimpedance-derived parameters track increasing alcohol levels~\cite{czaplik2019evaluation}.

Wearable and mobile inertial sensing has also been used to estimate intoxication outside of the driving context, typically by leveraging gait- and motion-derived features from commodity accelerometers or gyroscopes. For example, \NoHyper\citeauthor{chawathe2020using}\endNoHyper\ used smartphone accelerometer signals to classify intoxication/\ac{BAC}-related states from gait-derived features and reported discriminative performance around \ac{AUROC} $\approx 0.85$~\cite{chawathe2020using}. In a field setting, \NoHyper\citeauthor{killian2019learning}\endNoHyper\ collected smartphone accelerometer data during a one-day event with 13 students and used an ankle-worn transdermal alcohol sensor to provide objective ground-truth labels; their best-performing classifier achieved 77.5\% accuracy for detecting intoxicated versus sober windows ($\ge$  \gdL{0.08} vs.\ $<$ \gdL{0.08}) on 10-second segments~\cite{killian2019learning}. \NoHyper\citeauthor{mcafee2017alcowear}\endNoHyper\ proposed AlcoWear, which infers intoxication from gait using inertial features extracted from smartphone and smartwatch accelerometer and gyroscope sensors, reporting 79.8\% accuracy using smartwatch data and 89.5\% using smartphone data for classifying \ac{BAC} ranges~\cite{mcafee2017alcowear}. More recently, \NoHyper\citeauthor{segura2025advancing}\endNoHyper\ collected smartwatch inertial signals (accelerometer and gyroscope) together with \ac{HR} over three weeks from 14 participants, using transdermal alcohol concentration as ground truth, and achieved an \ac{AUROC} of 0.75 for detecting intoxication above \ac{BAC} \gdL{0.05} using time-series \ac{ML} models~\cite{segura2025advancing}.

Taken together, prior wearable-based alcohol detection research demonstrates that alcohol-related signals can be captured using body-worn sensors, but the existing evidence stems predominantly from non-driving settings, specialized sensing modalities, or movement patterns such as gait. In contrast, our study investigates whether off-the-shelf smartwatches can detect alcohol-impaired driving in a real vehicle using only physiological and motion signals in a driving context.

\subsection{In-Vehicle Vital Signs and Drunk-Driving Detection}
\label{secC:in_vehicle_vital_signs_and_drunk_driving_detection}

Automotive environments are increasingly equipped with sensors capable of monitoring driver vital signs. Prior work has demonstrated \ac{ECG} acquisition from the steering wheel \cite{hamza2020monitoring} and hybrid sensing solutions combining \ac{ECG}, \ac{PPG}, and camera-based monitoring \cite{warnecke2023robust}. Remote \ac{PPG} extraction from in-cabin cameras has also advanced significantly \cite{huang2020heart, bouraffa2025deep}. Vital signs have also been measured using seat-integrated sensors in vehicles \cite{sidikova2020vital}. Commercial offerings integrating such capabilities are emerging, including from BMW \cite{BMWVitalSigns2025} and Bosch \cite{BoschInteriorsSensing}.

Prior studies have investigated drunk-driving detection using in-vehicle sensing and driver monitoring to infer impairment from deviations in driving behavior. Early work primarily relied on vehicle-state and driver-control signals, and many studies relied on simulator-based experiments to classify intoxication from steering patterns, lane positioning, and other behavioral metrics \cite{li2015drunk, sun2018recognition}. More recent work typically combines high-resolution vehicle telemetry (e.g., steering, lane position, pedal inputs) with \acp{DMC} capturing eye movements, gaze stability, and head pose, analyzed using \ac{ML} to distinguish intoxicated from sober driving in simulators~\cite{lee2010assessing, chen2017support, li2015drunk, li2020random, koch2023leveraging}. Extending this line of work to real-vehicle settings, a recent test-track study combining driver-vehicle interaction and \ac{DMC} data reported an \ac{AUROC} of 0.84 \cite{Deuber2025}. Complementary evidence from non-detection-oriented studies shows that camera-based indicators of impaired oculomotor control and disrupted visual scanning reliably reflect alcohol intoxication in both controlled and on-road driving contexts~\cite{maurage2020eye, zemblys2024practical, ahlstrom2023alcohol}. Despite high discriminative performance, this line of work relies on additional in-vehicle sensing hardware, proprietary vehicle integration, and careful calibration, which constrains accessibility due to hardware costs and limits applicability largely to recent vehicle generations and research-grade settings~\cite{ANPRM2023, SmartEyeVolvoDMC2023}. Accordingly, our work complements this line of research by shifting the sensing locus from the vehicle to the driver: rather than relying on onboard telemetry, cameras, or proprietary vehicle integration, we study whether alcohol-impaired driving can be detected using only off-the-shelf smartwatch signals.

\subsection{Research Gap}
\label{secC:summary_and_research_gap}

Wearable and mobile intoxication detection has primarily been studied in non-driving contexts, often relying on gait and general movement patterns. Only recent work has begun to combine motion- and physiological-signal–based detection. Driving, however, is a highly structured activity with constrained movements and reduced variability compared with general human motion. This structure may facilitate classification, but many informative cues available in unconstrained settings, such as walking-related patterns, are absent. Because the underlying movement distribution differs substantially, it remains unclear whether motion-based approaches transfer to driving contexts. In addition, some wearable-based approaches rely on sensors that are not available in consumer smartwatches, such as transdermal alcohol concentration sensors. It therefore remains unclear whether off-the-shelf smartwatches can detect alcohol-impaired driving. Beyond smartwatches, some wearable approaches, such as fingertip \ac{PPG} sensors, are not feasible in driving contexts, while many other wearable-based intoxication detection systems have so far been evaluated only in laboratory settings. In contrast to these approaches, in-vehicle intoxication detection predominantly focuses on driving behavior and onboard sensing. As a result, wearable-based intoxication detection in driving contexts remains underexplored. This gap matters because detection approaches that depend on vehicle-integrated sensing or proprietary telemetry cannot scale beyond newer vehicles, whereas wearable-based methods can be deployed independently of vehicle infrastructure. To the best of our knowledge, no prior work has investigated drunk-driving detection using wearable physiological and accelerometer signals in real driving scenarios.

\section{Data Collection}
\label{secC:data}

We conducted a randomized, controlled study (ClinicalTrials.gov NCT05796609) to collect physiological and behavioral data from drivers across varying levels of alcohol intoxication as they were driving. Data collection took place on a closed test track to maximize safety while closely approximating real-world driving conditions. The study protocol was approved by the local ethics committee (IRB) in Bern, Switzerland (ID~2022-02245). It was conducted between April 2023 and July 2023.

\subsection{Participants}

We recruited 72 participants through public advertisements and social media. Of these, 55 met the inclusion criteria, and 1 was excluded from the analysis due to missing data, resulting in a final sample of $n = 54$. Eligible individuals were at least 21 years old, held a valid driving license for a minimum of three years, and reported regular driving activity. We excluded individuals with a history of substance abuse, pregnancy, or medical conditions contraindicating alcohol consumption. To strictly control for habitual alcohol misuse, we employed a two-stage screening process. First, candidates completed the \ac{AUDIT} \cite{saunders1993development}, and those with scores of 15 or higher were excluded. Second, eligible candidates attended a physical screening visit where capillary blood samples were analyzed for \ac{PEth}, an objective biomarker of alcohol consumption \cite{luginbuhl2022consensus, schrock2017phosphatidylethanol}. Candidates with \ac{PEth} levels indicating excessive chronic consumption (greater than 200~ng/mL) were excluded. The resulting sample consisted of 54 participants (age \meanstd{37.2}{14.9}~years), with a balanced gender distribution (28 female, 26 male). All participants provided written informed consent prior to the study and received financial compensation for their participation (200~CHF, equivalent to 222~USD).

\subsection{Study Design}

The study employed a three-group design to isolate the effects of alcohol from potential confounding factors such as fatigue, learning effects, and placebo responses. 
We assigned 31 participants to the treatment group, and they received alcoholic beverages tailored to induce specific \ac{BAC} trajectories. They completed driving sessions at three intoxication levels: a sober baseline, a severe intoxication phase (target peak \ac{BAC} \gdL{0.08}; complete session above \gdL{0.05}), and a moderate intoxication phase on the descending limb of the alcohol curve (below \gdL{0.05}, i.e., below the \ac{WHO}-recommended legal limit). Treatment participants were blinded to both alcohol presence and dosage. Another 12 participants were assigned to the placebo group \cite{testa2006understanding}. To control for expectancy effects associated with consuming a beverage believed to contain alcohol, these participants received a non-alcoholic placebo drink that mimicked the appearance of the treatment beverage and were blinded to their group assignment. The remaining 11 participants were assigned to the open-label reference group. This group received no beverages and completed all driving sessions in a sober state. Participants in the reference group were fully informed about their condition. This group served to control for circadian influences (e.g., increasing drowsiness over the course of the day) and practice effects related to repeated driving of the test track. We selected group sizes consistent with those commonly used in the driver state detection literature (in particular, \cite{koch2023leveraging, lehmann2024machine}).

\subsection{Apparatus and Sensors}

We collected physiological signals and accelerometer data using a consumer-grade smartwatch (Garmin~vivoactive~4S) that participants wore on their right wrist. Participants fitted the smartwatch securely, and the study team then verified the fit and adjusted it if needed to minimize motion artifacts during steering maneuvers while maintaining comfort. The smartwatch continuously recorded \acp{IBI} in milliseconds (derived from \ac{PPG}), \ac{HR} in beats per minute (based on \ac{PPG}), and triaxial accelerometer data in $g$.

Breath alcohol concentration served as the ground truth for intoxication levels. Measurements were taken using a police-grade breathalyzer (Dräger~6820, Drägerwerk AG \& Co. KGaA), which is certified for law-enforcement use. Breath alcohol concentration values were converted to \ac{BAC} using a standard breath-to-blood conversion factor of 0.2 (i.e., \mgL{0.25} breath alcohol concentration corresponds to \gdL{0.05} \ac{BAC}).

\subsection{Procedure} 

\begin{figure}[htbp]
  \centering
  \includegraphics[width=\linewidth]{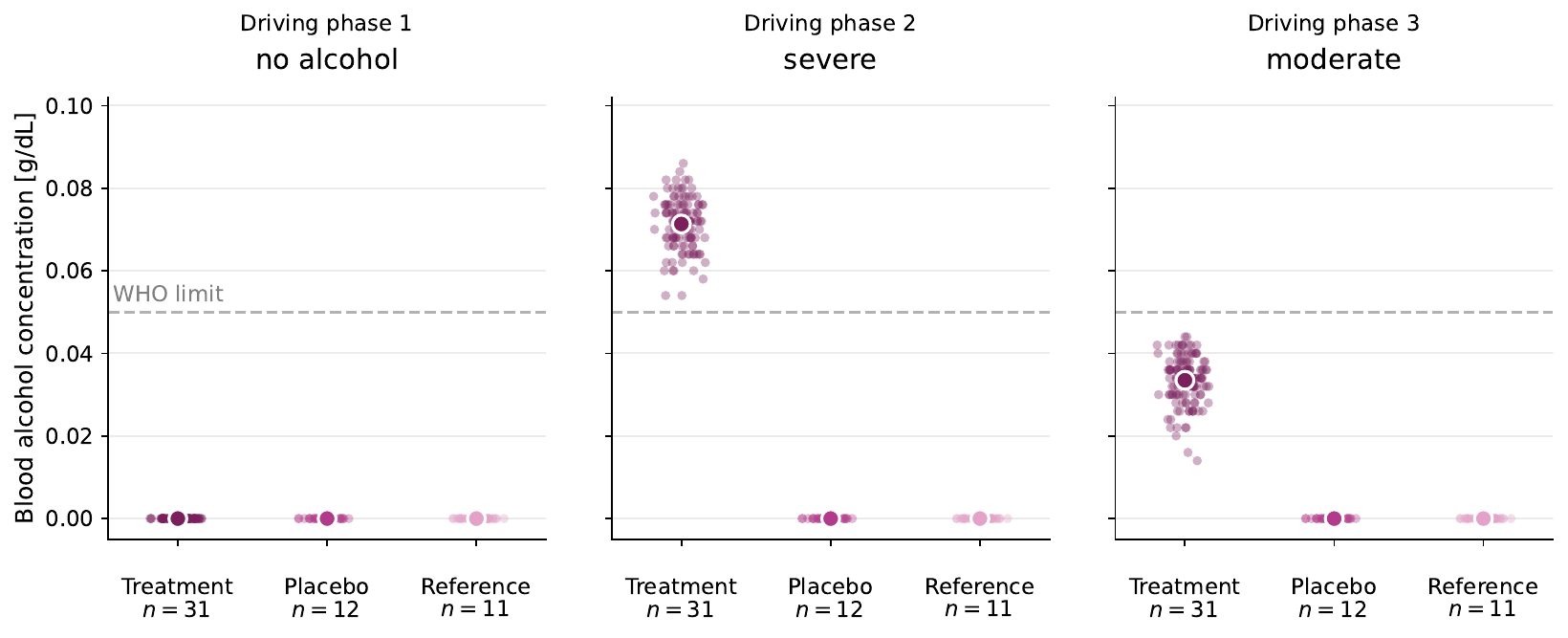}
  \caption{Blood alcohol concentration across driving phases. Mean blood alcohol concentration (BAC; \gdL{}) for the treatment (n=31), placebo (n=12), and reference (n=11) groups shown separately for driving phases 1–3. Points indicate group means across participants. The dashed horizontal line marks the WHO-recommended BAC limit (\gdL{0.05}).}
  \Description{} 
  ~\label{figC:BACs}
\end{figure}

Each study day corresponded to the duration of a full workday for a participant. Participants were requested to arrive in a fasted state (no food intake within four hours prior to arrival) to ensure comparable metabolic conditions. Upon arrival, participants underwent a urine test to screen for other substances and, where applicable, pregnancy. 

Participants first completed a familiarization drive to acclimatize to the vehicle handling and the test-track layout. Subsequently, all participants, independent of group assignment, completed a baseline driving session in a confirmed sober state (\ac{BAC} = \gdL{0.00}). After the baseline drive, the alcohol administration phase commenced for the treatment group. Individual alcohol doses were calculated using the Widmark formula \cite{widmark1932}, adjusting for age, sex, body weight, and height to set a dose-calculation target \ac{BAC} of \gdL{0.08}. The alcoholic beverage was vodka mixed with bitter orange juice to mask its taste. To maintain blinding, the placebo group received an identical volume of bitter orange juice as placebo drink. Both groups received their beverages in neutral bottles with narrow outlets to reduce exposure to smell. Beverages were consumed in a controlled environment over a fixed 20-minute period. Usually, only one participant was present during administration; if two participants were present, they were instructed not to discuss their perceived alcohol level or the drinks they received.

A visualization of participant \acp{BAC} is provided in Figure~\ref{figC:BACs}. For the treatment group, the second driving session (\emph{severe} phase) typically began once \ac{BAC} peaked and subsequently fell below \gdL{0.075}. Due to procedural variability (e.g., measurement fluctuations), some participants partially drove above this threshold. However, all observed values remained between \gdL{0.054} and \gdL{0.086}. A waiting period of at least 20 minutes after the last sip was enforced to prevent mouth-alcohol contamination of breathalyzer measurements. After the severe phase drive, participants rested while their \ac{BAC} naturally decreased. The third driving session (\emph{moderate} phase) was initiated when \ac{BAC} generally had descended below \gdL{0.035}, with observed values ranging from \gdL{0.014} to \gdL{0.044}. Breathalyzer tests were administered immediately before and after each driving scenario to obtain precise ground truth labels for all data segments. Participants in the placebo and reference groups followed the same temporal schedule as the treatment group. Their second and third drives were scheduled at time points matched to the treatment group to ensure comparable fatigue and circadian states across all conditions.

\begin{figure}[htbp]
  \centering
  \includegraphics[width=\linewidth]{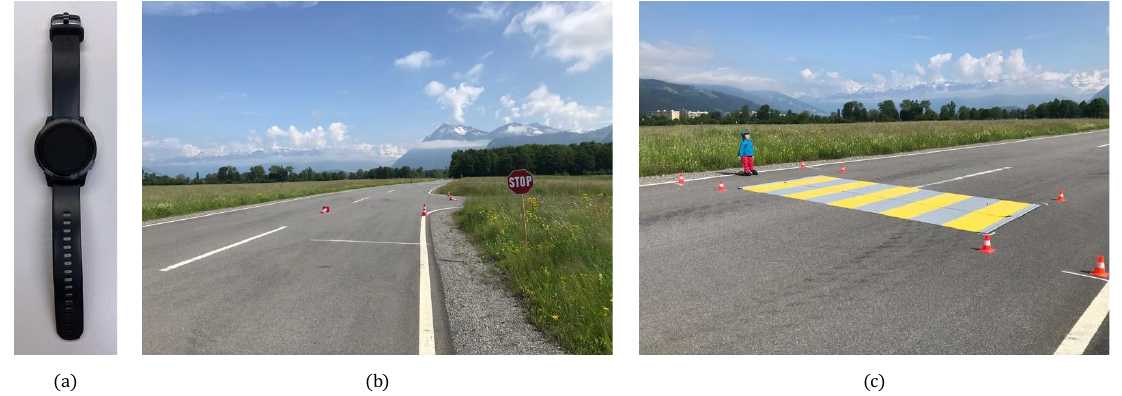}
  \caption{Illustrations of the experimental setup: (a) off-the-shelf smartwatch worn by participants; (b) temporary crossroads on the test track where participants were required to stop; and (c) temporarily marked crossing area on the track delineated with cones.}
  \Description{} 
  ~\label{figC:study_images}
  \end{figure}

\subsection{Driving Tasks and Environment}

The driving tasks took place on a closed-circuit proving ground in Switzerland (see Figure \ref{figC:study_images}). The study vehicle was a VW~Touran with automatic transmission. The track featured a mix of road geometries, including long straight sections, wide curves, and a complex inner segment with tight turns and intersections. Road widths ranged from 6 to 10~meters. To approximate realistic driving demands, we designed three distinct scenarios. On average, each driving phase (i.e., one of the three) lasted 40 minutes. The \emph{highway} scenario involved high-speed driving (up to \kmh{80}) on the track's outer oval. The \emph{rural} scenario was a mixed-speed segment (up to \kmh{60}) that included moderate curves, stop signs, and obstacles to maneuver around. The \emph{urban} scenario was a lower-speed segment (up to \kmh{50}) involving tight turns, a pedestrian crosswalk requiring a full stop, and navigation around artificial obstacles (traffic cones representing roadworks or parked vehicles). Each driving session (no alcohol, severe, moderate) consisted of completing all three scenarios. To prevent order effects, the sequence of scenarios (e.g., urban--highway--rural) and the direction of travel were randomized for each participant and each session. A licensed driving instructor was present in the passenger seat at all times, equipped with dual control pedals to intervene in the event of safety-critical errors.

\section{Modeling and Evaluation}
\label{secC:methods}

\subsection{Data Preparation}

Two time series were extracted for each participant: a physiological arousal signal and an accelerometer-based motion signal. As a basis for the arousal signal, the smartwatch provides \ac{IBI} measurements, which are event-based and irregularly sampled, making direct application of standard time-series modeling techniques challenging. We therefore transform \ac{IBI} and heart-rate data into an equally spaced physiological representation using a pretrained arousal-estimation model. This choice is grounded in established physiological theory. Acute alcohol intake acts as a systemic stressor that activates the sympathetic nervous system and suppresses parasympathetic activity, leading to reduced heart rate variability and elevated heart rate, which are hallmarks of physiological arousal. Prior work has shown that alcohol-induced autonomic responses resemble those elicited by other physical and cognitive stressors. We leverage the arousal-estimation model developed by \NoHyper\citeauthor{mishra2020evaluating},\endNoHyper\ which was trained on diverse arousal paradigms including mental, startle-based, and physical stressors \cite{mishra2020evaluating}. This diversity enables robust characterization of autonomic activation across heterogeneous conditions. Our approach thus treats alcohol-induced impairment as a form of physiological arousal, grounded in shared autonomic mechanisms, while enabling stable and interpretable time-series modeling. However, this signal should not be interpreted as an alcohol-specific biomarker. Elevated physiological arousal can also arise from other sources of arousal, such as physical activity, task demands, and fatigue.
 
Before inferring physiological arousal probabilities, we removed outliers and applied participant-specific z-score normalization based on each participant's data distribution. After cleaning both streams, short-window features were computed for \ac{IBI} (mean, standard deviation, median, minimum, maximum, 20th percentile, 80th percentile, and root mean square of successive differences) and for \ac{HR} (mean, standard deviation, median, 20th percentile, and 80th percentile) and passed to the pretrained model (provided by \citet{mishra2020evaluating}) to obtain a continuous physiological arousal probability over time. Predicted arousal probabilities range from 0 to 1, with values closer to 0 indicating lower physiological arousal and values closer to 1 indicating higher physiological arousal. Accelerometer data was processed independently by loading tri-axial acceleration and computing a motion-intensity signal as the magnitude of the acceleration vector.

For the logistic regression approach (in contrast to the \ac{CNN} pipeline), we perform an additional feature-extraction step. From the two time series (physiological arousal and acceleration), we constructed windowed segments that served as the basic modeling units. A sliding window with a fixed length (180 s) and a step size (45 s) produced overlapping segments that captured short-term dynamics in both physiological arousal and motion signals. We selected this window size as a compromise between faster detection with smaller windows and improved detection performance with larger windows, consistent with the general window-length trade-off discussed by \NoHyper\citeauthor{Deuber2025}\endNoHyper\ for drunk-driving detection using in-vehicle sensors \cite{Deuber2025}. For each window, we extracted the corresponding sequences of arousal probability and motion intensity and, in addition, computed a broad set of statistical and temporal features using the tsfresh library \cite{christ2018time}, including distributional, autocorrelation, and complexity-based measures. Feature extraction was applied separately to the physiological arousal and accelerometer streams, and only windows with sufficient data coverage (> 50 \%) and complete overlap with driving intervals were retained. In total, we extracted 783 features per modality; Table~\ref{tabC:tsfresh_feature_families} summarizes the feature families.

\begin{table}[htbp]
\centering
\caption{Overview of extracted \texttt{tsfresh} features per modality.}
\label{tabC:tsfresh_feature_families}
\begin{tabular}{llr}
\toprule
feature family & \texttt{tsfresh} name & \# features \\
\midrule
frequency-domain / spectral features &
fft\_coefficient, fft\_aggregated,  &
422 \\
& fourier\_entropy, spkt\_welch\_density,  & \\
& energy\_ratio\_by\_chunks & \\
distribution / quantile features &
quantile, index\_mass\_quantile, & 77\\
& change\_quantiles, binned\_entropy &
 \\
wavelet / time--frequency features &
cwt\_coefficients, number\_cwt\_peaks &
62 \\
trend / regression features &
linear\_trend, agg\_linear\_trend &
53 \\
autocorrelation / dependence features &
autocorrelation, partial\_autocorrelation, & 23 \\
& agg\_autocorrelation & \\
counts / thresholding / crossings &
value\_count, range\_count, count\_above, & 21 \\
& count\_below, number\_crossing\_m, & \\
& ratio\_beyond\_r\_sigma & \\
entropy / complexity features &
approximate\_entropy, sample\_entropy, & 18 \\
& permutation\_entropy, & \\
& lempel\_ziv\_complexity, cid\_ce &
 \\
summary statistics &
mean, median, variance, & 17 \\
& standard\_deviation, skewness, kurtosis, & \\
& minimum, maximum, \ldots &
 \\
autoregressive features &
ar\_coefficient &
11 \\
nonlinear dynamics features &
friedrich\_coefficients, & 5 \\ 
& max\_langevin\_fixed\_point & \\
peak-related features &
number\_peaks &
5 \\
boolean / duplicate indicators &
has\_duplicate, has\_duplicate\_max, & 3 \\
&  has\_duplicate\_min & \\
stationarity / unit-root test features &
augmented\_dickey\_fuller &
3 \\
similarity / query-matching features &
query\_similarity\_count &
1 \\
other / uncategorized features & &
62 \\
\bottomrule
\end{tabular}
\end{table}

All samples were aligned with the study protocol by mapping timestamps to the three driving phases. Treatment participants completed a no alcohol phase (\emph{phase~1}), a severe intoxication phase (\ac{BAC} $>$ \gdL{0.05}, \emph{phase~2}), and a moderate intoxication phase starting once \ac{BAC} had typically fallen below \gdL{0.035} (\emph{phase~3}). Placebo and reference participants followed the same schedule but remained sober throughout. These phases defined the labels used in downstream modeling. We consider two binary detection tasks that reflect complementary goals: \EARLY{} captures any alcohol exposure (\ac{BAC} $>$ \gdL{0.00}) as an early-warning signal, whereas \ABOVE{} targets episodes above the \ac{WHO}-recommended limit (\ac{BAC} $>$ \gdL{0.05}). Figure~\ref{figC:pipeline} visualizes the two classification tasks.

For the \EARLY{} task, all windows with \ac{BAC} $>$ \gdL{0.00} (treatment phases~2 and~3) were labeled positive; for the \ABOVE{} task, only windows with \ac{BAC} $>$ \gdL{0.05} (treatment phase~2) were labeled positive. All placebo and reference samples were labeled negative. Beyond the \EARLY{} and \ABOVE{} binary tasks, we also considered categorical classification tasks trained separately for treatment and control participants aiming to distinguish driving phases~1--3, to evaluate whether the models captured alcohol-related impairment patterns rather than the effects attributable to circadian rhythms, fatigue, or cumulative physiological arousal over the course of the day, thus disambiguating potential ordering effects.

 \subsection{Models and Evaluation}
We evaluated two modeling approaches: a logistic regression model with LASSO regularization and a two-tower (late-fusion) 1D \ac{CNN} trained directly on windowed time-series inputs. Both models were applied to the two binary alcohol-detection tasks. In the categorical setting, models were trained separately for treatment and control participants (placebo and reference groups). All evaluations followed a \ac{LOSO} validation scheme, in which we iteratively hold out one participant for testing while training on the remaining participants. 

As a baseline approach, we implemented logistic regression with LASSO regularization. Our pipeline is conceptually similar to the in-vehicle sensor-based drunk-driving detection approach of \NoHyper\citeauthor{koch2023leveraging}\endNoHyper\, but differs in sensing modality and feature construction \cite{koch2023leveraging}. Specifically, our baseline uses a substantially larger feature set computed via the tsfresh package~\cite{christ2018time}. As an additional robustness analysis, we evaluated multiple window configurations, combining window sizes from 30~s to 600~s with step sizes equal to one quarter of the respective window length and a minimum window coverage of 50\% (i.e., at least 50\% of expected samples present). The model was trained with class-balanced weights using the liblinear solver. Probabilistic predictions for the held-out participant were used to assess the performance metrics as described above.

Beyond the feature-based logistic-regression baseline, we trained a two-tower 1D \ac{CNN}, operating directly on time-series segments from raw physiological arousal and accelerometer signals. This \ac{CNN} constitutes a step toward end-to-end learning, while still relying on the pretrained physiological-arousal-estimation model to transform irregular physiological measurements into an equally spaced input sequence. We use the \ac{CNN} model to examine whether learned temporal representations can capture impairment-related structure beyond what is available from hand-crafted features alone. We adopted a two-tower late-fusion architecture because the physiological arousal and accelerometer streams differ substantially in temporal resolution and signal characteristics (1 Hz vs. 25 Hz), including temporal granularity, smoothness, and likely noise structure. Moreover, they reflect complementary dimensions of impairment, namely autonomic arousal and motor behavior. Accordingly, separate modality-specific convolutional towers first encode each stream before their representations are fused in a shared classification head.

\begin{figure}[htbp]
  \centering
  \includegraphics[width=0.6\linewidth]{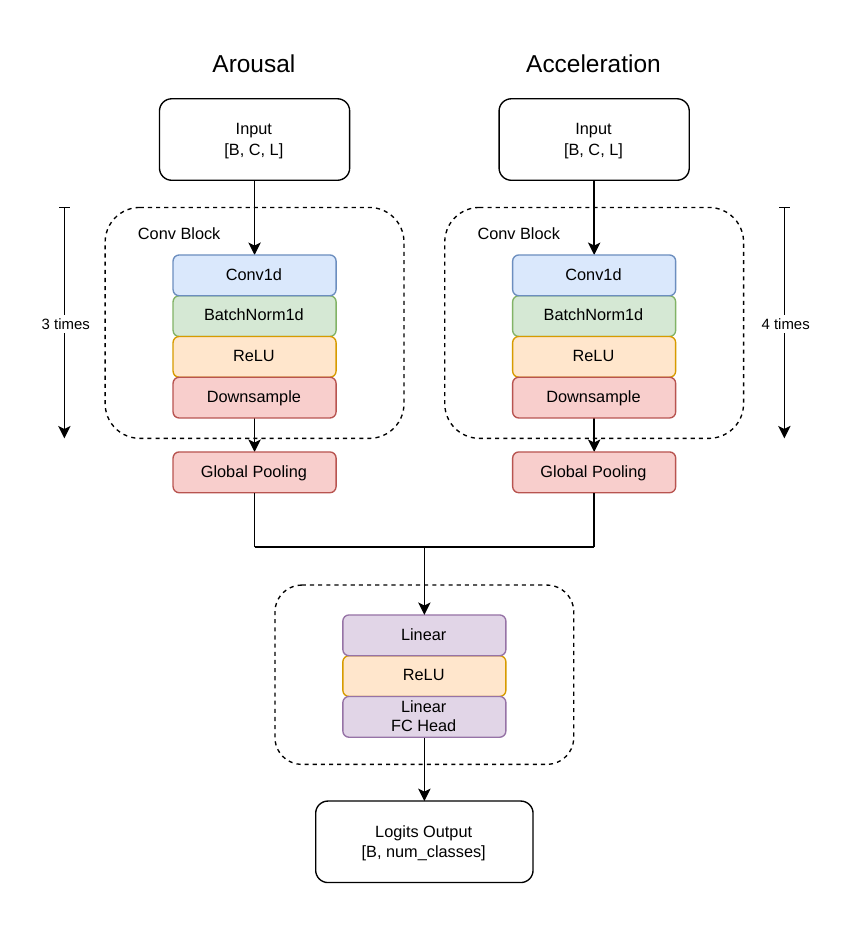}
  \caption{Two-tower (late-fusion) 1D \ac{CNN} architecture. Here, \emph{B} denotes batch size, \emph{C} channels, and \emph{L} window length.}
  \Description{TBD.} 
  ~\label{figC:architecture}
\end{figure}

To operationalize this architecture, both modalities were segmented into fixed-length windows (180 s) with a 15 s step, using the same window length as in the logistic regression approach. Within each window, both modalities were resampled to regular temporal grids: the physiological arousal signal was mapped to a 1 Hz grid, and the accelerometer magnitude to a 25 Hz grid (40 ms). Windows were retained only when at least one-third of the expected samples were present. We used a more permissive coverage threshold than in the logistic-regression pipeline because the 1D \ac{CNN} requires a fixed, exact number of input samples per window. Accordingly, missing values were imputed via time-based linear interpolation; any remaining leading or trailing gaps were filled via forward and backward fill. This procedure produced two synchronized sequences per window (one per modality) that were passed to the respective convolutional towers of the 1D \ac{CNN} (see Figure \ref{figC:architecture}). The physiological arousal tower of the \ac{CNN} comprises 3 repeated 1D convolution blocks (Conv1d + BatchNorm + ReLU with temporal downsampling with 16, 32, and 64 channels and kernel size 5), whereas the accelerometer-magnitude tower comprises 4 such blocks (32, 64, 128, and 128 channels and kernel size 7). Each tower applies adaptive global average pooling to obtain a fixed-length embedding; embeddings are concatenated and passed to a shared fully connected head that outputs logits, with dropout (rate 0.3) applied during training.

In each \ac{LOSO} fold, one participant was held out for testing, and the remaining participants were split into training and validation sets by randomly selecting 10 ($\approx$20\%) for validation. Before training, each modality was initially standardized using z-score normalization fitted on the inner training data. The model was optimized with AdamW, using a class-weighted binary cross-entropy loss to address imbalance, together with a ReduceLROnPlateau scheduler. For the categorical phase classification, we trained the models using categorical cross-entropy loss. Training employed early stopping based on validation \ac{AUROC}, and the best-performing model per fold was subsequently evaluated on the held-out participant. As in the logistic regression baseline, macro- and micro-level \ac{AUROC} and \ac{AUPRC} were computed for comparability across experiments. 

As an additional test, we report \ac{CNN} results under a per-phase normalization scheme. Specifically, we compute z-scores within each participant and phase and apply them to the corresponding phase data. This per-phase normalization represents a deliberately conservative evaluation setting. By independently z-normalizing each participant’s signals within each driving phase, we explicitly remove phase-level mean and variance shifts, setting each phase to zero mean and unit variance. This procedure suppresses baseline differences between sober and intoxicated driving that may arise from alcohol-induced physiological changes (e.g., elevated heart rate and reduced \ac{HRV}). As a result, the model must rely primarily on finer-grained temporal and structural patterns in physiological and motion signals, rather than global level shifts. While this normalization likely removes informative signal and is therefore expected to reduce performance, it provides a stringent test of whether discrimination is driven by dynamic impairment-related patterns rather than coarse baseline differences.

As an additional analysis, we evaluated whether the proposed model can predict continuous \ac{BAC} values rather than solving binary classification tasks. While \gdL{0.05} is the \ac{WHO}-recommended limit and one of the most commonly applied operational thresholds \cite{who2024alcohol}, \ac{BAC} thresholds vary across jurisdictions~\cite{DrinkDrivingLimitsWW}. Modeling \ac{BAC} as a continuous outcome therefore provides a more comprehensive view of performance across the full intoxication spectrum. We used the same two-tower \ac{CNN} architecture, input modalities, training procedure, and \ac{LOSO} evaluation framework as in the classification setting. The main differences concerned target construction, model output, loss function, and evaluation. Specifically, instead of assigning each time window a binary label, we derived a continuous BAC target for each timestamp by linearly interpolating between time-stamped \ac{BAC} measurements. The final output layer was adapted to produce a single continuous prediction rather than a binary logit, and the model was optimized using smooth L1 loss. We summarize regression performance using mean absolute error and Pearson correlation between predicted and reference BAC values. To maintain comparability with the \ABOVE{} classification task, we additionally report the \ac{AUROC} obtained when thresholding the regression outputs to identify whether participants were above \gdL{0.05}.

\section{Results}
\label{secC:results}

We report performance for two binary classification tasks: \EARLY{} and \ABOVE{}.

Only participants in the treatment group exhibit both positive and negative labels, whereas placebo and reference participants contain exclusively negative samples. Consequently, per-participant \ac{AUROC}/\ac{AUPRC} can only be computed for treatment participants, and macro-averages (per-participant averages; \meanstd{mean}{std} across \ac{LOSO}-held-out participants) therefore characterize participant-level discrimination within this group. In contrast, to evaluate performance in a more deployment-relevant setting that includes both intoxicated and consistently sober drivers, we additionally report pooled predictions that incorporate control participants (micro-averages). Because these participants lack positive labels, a pooled evaluation is required to include them. For completeness and comparability, we also report pooled predictions for the treatment group alone. To contextualize \ac{AUPRC} under varying class imbalance, we report a chance-level baseline defined by the positive-class prevalence of the respective evaluation split: in precision–recall space, an uninformative classifier achieves an \ac{AUPRC} equal to this prevalence~\cite{saito2015precision}.

\subsection{Logistic Regression Model}

\begin{table*}
     \centering
    \Description{TBD} 
    \caption{Performance of LASSO-regularized logistic regression (180 s windows) for the \EARLY{} and \ABOVE{} binary alcohol-detection tasks; as well as ablations of physiological-arousal-only and accelerometer-only (acc). We report macro-averaged (per-participant) and micro-averaged (pooled) \ac{AUROC} and \ac{AUPRC}, together with random \ac{AUPRC} baselines, for treatment participants and for the combined treatment+control group.}
    \label{tabC:results_LR}
    \begin{tabular}{cllcccc}
        \toprule
        & & & \multicolumn{2}{c}{\EARLY} & \multicolumn{2}{c}{\ABOVE} \\ 
        \toprule \noalign{\vskip 10pt}
        \multirow{3}{*}{\shortstack[c]{per-participant\\average}} & \multirow{3}{*}{treatment} & \ac{AUROC}  & \multicolumn{2}{c}{\meanstd{0.80}{0.11}} &  \multicolumn{2}{c}{\meanstd{0.75}{0.10}}  \\ 
         & & \ac{AUPRC}   & \multicolumn{2}{c}{\meanstd{0.89}{0.08}} &  \multicolumn{2}{c}{\meanstd{0.60}{0.13}}  \\ 
         & & \random{rand. \ac{AUPRC}}
         &  \multicolumn{2}{c}{\random{\meanstd{0.67}{0.03}}} & \multicolumn{2}{c}{\random{\meanstd{0.33}{0.02}}}  \\
         \hline
        \multirow{6}{*}{\shortstack[c]{pooled\\predictions}} & \multirow{3}{*}{treatment} & \ac{AUROC}   & \multicolumn{2}{c}{0.77} &  \multicolumn{2}{c}{0.71}  \\
         & & \ac{AUPRC}   & \multicolumn{2}{c}{0.87} & \multicolumn{2}{c}{0.58}  \\ 
         & & \random{rand. \ac{AUPRC}} & \multicolumn{2}{c}{\random{0.67}} & \multicolumn{2}{c}{\random{0.34}} \\
         \cline{2-7} 
         & \multirow{3}{*}{\shortstack[c]{treatment\\+ control}} & \ac{AUROC}   & \multicolumn{2}{c}{0.73} & \multicolumn{2}{c}{0.72}   \\
         & & \ac{AUPRC}   & \multicolumn{2}{c}{0.61} &  \multicolumn{2}{c}{0.40}  \\
         & & \random{rand. \ac{AUPRC}} & \multicolumn{2}{c}{\random{0.38}} & \multicolumn{2}{c}{\random{0.19}} \\
      \hline \noalign{\vskip 15pt}
     & & & arousal only & acc only & arousal only & acc only \\[2pt]
     \cline{4-7}\noalign{\vskip 2pt}
        \multirow{3}{*}{\shortstack[c]{per-participant\\average}} & \multirow{3}{*}{treatment} & \ac{AUROC}  & \meanstd{0.70}{0.14} & \meanstd{0.74}{0.14} & \meanstd{0.64}{0.12} & \meanstd{0.73}{0.11} \\ 
         & & \ac{AUPRC}   & \meanstd{0.83}{0.10} & \meanstd{0.84}{0.10} & \meanstd{0.46}{0.13} & \meanstd{0.59}{0.15} \\ 
         
         & & \random{rand. \ac{AUPRC}}
      & \random{\meanstd{0.67}{0.03}} & \random{\meanstd{0.67}{0.03}} & \random{\meanstd{0.33}{0.02}} & \random{\meanstd{0.33}{0.02}} \\
      \hline
        \multirow{6}{*}{\shortstack[c]{pooled\\predictions}} & \multirow{3}{*}{treatment} & \ac{AUROC}  & 0.70 & 0.70 & 0.64 & 0.69 \\
         & & \ac{AUPRC}   & 0.83 & 0.81 & 0.44 & 0.55 \\ 
         & & \random{rand. \ac{AUPRC}}
      & \random{0.67} & \random{0.67} & \random{0.34} & \random{0.34} \\
      \cline{2-7} 
         & \multirow{3}{*}{\shortstack[c]{treatment\\+ control}} & \ac{AUROC}   & 0.62 & 0.70 & 0.61 & 0.72 \\
         & & \ac{AUPRC}   & 0.50 & 0.57 & 0.24 & 0.39 \\
         & & \random{rand. \ac{AUPRC}}
      & \random{0.38} & \random{0.38} & \random{0.19} & \random{0.19} \\
         \hline
    \end{tabular}
\end{table*}

\begin{figure}[!htbp]
  \centering
  \includegraphics[width=0.8\linewidth]{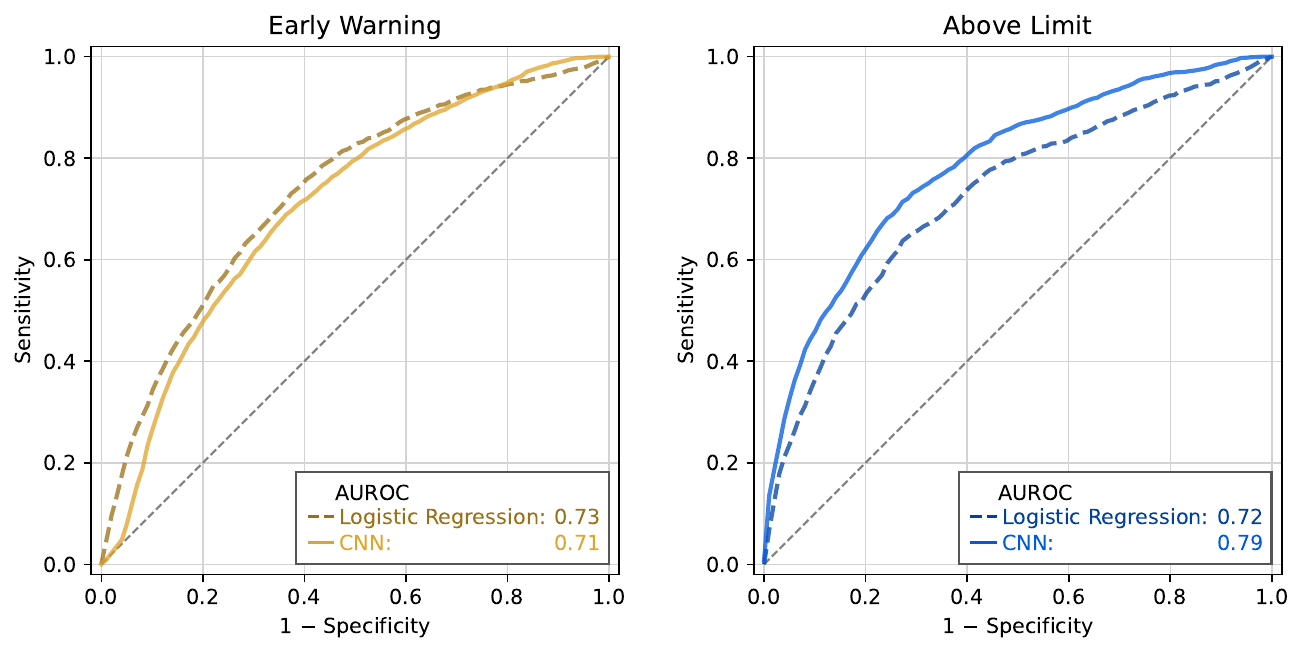}
    \caption{Receiver operating characteristic (ROC) curves for the \EARLY{} and \ABOVE{} tasks, comparing logistic regression and \ac{CNN} models on pooled predictions with treatment and control groups included. The dashed gray line indicates the performance of a random classifier.} 
  \Description{TBD.} 
  ~\label{figC:plot_ROC}
\end{figure}

Table~\ref{tabC:results_LR} summarizes the logistic regression baseline. Within the treatment group, the model achieved solid discrimination for \EARLY{} (\ac{AUROC} \meanstd{0.80}{0.11}; \ac{AUPRC} \meanstd{0.89}{0.08}), while performance was slightly lower for \ABOVE{} (\ac{AUROC} \meanstd{0.75}{0.10}; \ac{AUPRC} \meanstd{0.60}{0.13}). The \ac{AUPRC} gap is consistent with the stronger class imbalance in \ABOVE{} (random \ac{AUPRC} $\approx 0.33$ vs.\ $\approx 0.67$ for \EARLY{} in the treatment-only evaluation). Across both tasks, pooled (micro) performance was slightly lower than the corresponding per-participant averages (e.g., \EARLY{}: pooled \ac{AUROC} 0.77 vs.\ macro \ac{AUROC} \meanstd{0.80}{0.11}; \ABOVE{}: pooled \ac{AUROC} 0.71 vs.\ macro \ac{AUROC} \meanstd{0.75}{0.10}). When pooling treatment and control participants, \ac{AUPRC} decreased substantially, reflecting the lower prevalence of positive labels in that combined evaluation (random \ac{AUPRC} 0.38 for \EARLY{} and 0.19 for \ABOVE{}), whereas pooled \ac{AUROC} remained comparatively stable (0.73 and 0.72, respectively). Figure~\ref{figC:plot_ROC} shows the \ac{ROC} curves for the logistic regression models.

Table~\ref{tabC:results_LR} also compares models trained on physiological-arousal-only vs.\ accelerometer-only features. Using a single modality (and keeping the standard window size of 180 s) reduced performance relative to the combined model (Table~\ref{tabC:results_LR}), but both modalities retained measurable predictive value. In the treatment-only evaluation for \EARLY{}, physiological-arousal-only and accelerometer-only models achieved similar \ac{AUPRC} (both around \meanstd{0.83}{0.10}--\meanstd{0.84}{0.10}), while the accelerometer-only model yielded higher \ac{AUROC} than the physiological-arousal-only model. For \ABOVE{}, accelerometer-only clearly outperformed physiological-arousal-only (macro \ac{AUROC} \meanstd{0.73}{0.11} vs.\ \meanstd{0.64}{0.12}; macro \ac{AUPRC} \meanstd{0.59}{0.15} vs.\ \meanstd{0.46}{0.13}).

\begin{table*}[!htbp]
    \centering
    \Description{TBD} 
    \caption{Effect of window length on logistic regression performance for the \EARLY{} and \ABOVE{} tasks. Pooled \ac{AUROC} and \ac{AUPRC} are shown for treatment-only and treatment+control evaluations across window lengths from 30\,s to 600\,s, along with random \ac{AUPRC} baselines. The 180\,s window (marked with $^{\ast}$) corresponds to the default configuration used in subsequent analyses.}
    \label{tabC:results_LR_window-size}
    \begin{tabular}{cllccccccc}
    \toprule
    & \multicolumn{1}{l}{} & & 30 & 60 & 120 & 180$^{\ast}$ & 300 & 450 & 600 \\
    \midrule
    \multirow{6}{*}{\shortstack[c]{\textsc{Early}\\ \textsc{Warning}}} & \multirow{3}{*}{treatment} & \ac{AUROC}
        & 0.73 & 0.75 & 0.76 & 0.77 & 0.77 & 0.80 & 0.80 \\
     & & \ac{AUPRC}
        & 0.84 & 0.85 & 0.86 & 0.87 & 0.87 & 0.88 & 0.89 \\
     & & \random{rand. \ac{AUPRC}}
        & \random{0.67} & \random{0.67} & \random{0.67} & \random{0.67} & \random{0.67} & \random{0.67} & \random{0.67} \\ 
     \cline{2-10}
     & \multirow{3}{*}{\shortstack[c]{treatment\\+ control}} & \ac{AUROC}
        & 0.73 & 0.73 & 0.73 & 0.73 & 0.72 & 0.76 & 0.74 \\
     & & \ac{AUPRC}
        & 0.60 & 0.61 & 0.60 & 0.61 & 0.61 & 0.65 & 0.64 \\
    & & \random{rand. \ac{AUPRC}}
        & \random{0.38} & \random{0.38} & \random{0.38} & \random{0.38} & \random{0.38} & \random{0.38} & \random{0.38} \\ 
        \midrule
    \multirow{6}{*}{\shortstack[c]{\textsc{Above}\\ \textsc{Limit}}} & \multirow{3}{*}{treatment} & \ac{AUROC}
        & 0.67 & 0.70 & 0.71 & 0.71 & 0.70 & 0.71 & 0.71 \\
     & & \ac{AUPRC}
        & 0.51 & 0.55 & 0.57 & 0.58 & 0.55 & 0.56 & 0.58 \\
    & & \random{rand. \ac{AUPRC}}
        & \random{ 0.33} & \random{0.33} & \random{0.33} & \random{0.34} & \random{0.33} & \random{0.33} & \random{0.33} \\ 
      \cline{2-10}
     & \multirow{3}{*}{\shortstack[c]{treatment\\+ control}} & \ac{AUROC}
        & 0.71 & 0.72 & 0.73 & 0.72 & 0.72 & 0.72 & 0.71 \\
     & & \ac{AUPRC}
        & 0.36 & 0.40 & 0.41 & 0.40 & 0.39 & 0.39 & 0.39 \\
     & & \random{rand. \ac{AUPRC}}
        & \random{0.19} & \random{ 0.19} & \random{ 0.19} & \random{ 0.19} & \random{ 0.19} & \random{ 0.19} & \random{ 0.19} \\ 
     \hline
\end{tabular}
\end{table*}

Table~\ref{tabC:results_LR_window-size} analyzes the impact of window size. For \EARLY{} (treatment-only pooled evaluation), longer aggregation windows consistently improved performance: \ac{AUROC} increased monotonically from 0.73 (30 s) to 0.80 (600 s), and \ac{AUPRC} increased from 0.84 to 0.89. When pooling both treatment and control groups, the influence of window size was rather smaller and notable particularly in AUPRC, which increased from 0.60 (30 s) to 0.64 (600 s). For \ABOVE{}, the effect of window size was also limited: \ac{AUROC} remained within a relatively narrow range (0.67--0.71), and \ac{AUPRC} varied only modestly (0.51--0.58). A similar pattern was observed when pooling treatment and control participants, albeit at different \ac{AUPRC} levels.

Table~\ref{tabC:feature_family_coefficients_split} summarizes the mean absolute coefficients of the two logistic regression models across tsfresh feature families, together with the overall mean absolute coefficient values. Coefficients were computed within the same \ac{LOSO} evaluation framework and then averaged across folds. Overall, acceleration features received substantially larger coefficients than physiological arousal features in both classification tasks (0.409 vs.\ 0.109 for \EARLY{} and 0.454 vs.\ 0.186 for \ABOVE{}), indicating that the logistic regression models relied more strongly on acceleration-derived features. Within the acceleration modality, entropy / complexity features showed the largest mean absolute coefficients for both tasks. For \ABOVE{}, peak-related features ranked second, whereas the pattern was more mixed for \EARLY{}. Within the physiological arousal modality, the largest coefficients in \EARLY{} were observed for peak-related and entropy / complexity features. For \ABOVE{}, the most prominent physiological arousal feature families were entropy / complexity, peak-related, autoregressive, and trend / regression features.

\begin{table*}[t]
  \centering
  \caption{Mean absolute coefficients by tsfresh feature family, presented separately by classification task and modality. Values are reported as \meanstd{mean}{standard deviation}. The bottom row shows the overall mean and standard deviation across feature families. Missing entries denote feature families that were excluded because their features contained too many missing values after tsfresh extraction (specifically, 100.00\% missing values for similarity / query-matching features and 3.07\% missing values for nonlinear dynamics features in the physiological arousal modality).}
  \label{tabC:feature_family_coefficients_split}
  \small
  \begin{tabular}{lcccc}
  \toprule
  & \multicolumn{2}{c}{\EARLY{}} & \multicolumn{2}{c}{\ABOVE{}} \\
  \cmidrule(lr){2-3} \cmidrule(lr){4-5}
  feature family & arousal & acc & arousal & acc \\
  \midrule
  frequency-domain / spectral features &
  \meanstd{0.069}{0.002} &
  \meanstd{0.065}{0.001} &
  \meanstd{0.125}{0.004} &
  \meanstd{0.087}{0.003} \\
  distribution / quantile features &
  \meanstd{0.146}{0.006} &
  \meanstd{0.559}{0.031} &
  \meanstd{0.244}{0.011} &
  \meanstd{0.364}{0.021} \\
  wavelet / time--frequency features &
  \meanstd{0.045}{0.006} &
  \meanstd{0.024}{0.002} &
  \meanstd{0.043}{0.005} &
  \meanstd{0.049}{0.003} \\
  trend / regression features &
  \meanstd{0.145}{0.011} &
  \meanstd{0.540}{0.031} &
  \meanstd{0.354}{0.023} &
  \meanstd{0.604}{0.041} \\
  autocorrelation / dependence features &
  \meanstd{0.104}{0.006} &
  \meanstd{0.604}{0.048} &
  \meanstd{0.209}{0.020} &
  \meanstd{0.589}{0.034} \\
  counts / thresholding / crossings &
  \meanstd{0.035}{0.003} &
  \meanstd{0.147}{0.007} &
  \meanstd{0.036}{0.007} &
  \meanstd{0.132}{0.015} \\
  entropy / complexity features &
  \meanstd{0.172}{0.021} &
  \meanstd{0.976}{0.065} &
  \meanstd{0.362}{0.028} &
  \meanstd{1.806}{0.094} \\
  summary statistics &
  \meanstd{0.072}{0.010} &
  \meanstd{0.619}{0.066} &
  \meanstd{0.215}{0.022} &
  \meanstd{0.492}{0.047} \\
  autoregressive features &
  \meanstd{0.141}{0.010} &
  \meanstd{0.534}{0.052} &
  \meanstd{0.303}{0.017} &
  \meanstd{0.517}{0.054} \\
  nonlinear dynamics features &
  -- &
  \meanstd{0.599}{0.131} &
  -- &
  \meanstd{0.137}{0.124} \\
  peak-related features &
  \meanstd{0.191}{0.012} &
  \meanstd{0.589}{0.077} &
  \meanstd{0.302}{0.025} &
  \meanstd{1.149}{0.111} \\
  boolean / duplicate indicators &
  \meanstd{0.065}{0.009} &
  \meanstd{0.042}{0.006} &
  \meanstd{0.099}{0.010} &
  \meanstd{0.040}{0.007} \\
  stationarity / unit-root test features &
  \meanstd{0.135}{0.021} &
  \meanstd{0.274}{0.034} &
  \meanstd{0.064}{0.014} &
  \meanstd{0.259}{0.058} \\
  similarity / query-matching features &
  -- & -- & -- & -- \\
  other / uncategorized features &
  \meanstd{0.093}{0.007} &
  \meanstd{0.147}{0.011} &
  \meanstd{0.069}{0.007} &
  \meanstd{0.123}{0.011} \\
  \cmidrule(lr){2-3} \cmidrule(lr){4-5}
  overall mean and standard deviation & 
  \meanstd{0.109}{0.010} &
  \meanstd{0.409}{0.040} &
  \meanstd{0.186}{0.015} &
  \meanstd{0.454}{0.045} \\
  \bottomrule
  \end{tabular}
\end{table*}

\subsection{1D \ac{CNN} Model}

The \ac{CNN} preprocessing pipeline yielded 14{,}433 samples. Table~\ref{tabC:results_CNN} reports the \ac{CNN} results. For treatment participants, the \ac{CNN} improved per-participant average performance over logistic regression for both tasks, reaching \ac{AUROC} \meanstd{0.88}{0.09} and \ac{AUPRC} \meanstd{0.93}{0.05} for \EARLY{}, and \ac{AUROC} \meanstd{0.86}{0.11} and \ac{AUPRC} \meanstd{0.78}{0.17} for \ABOVE{}. At the pooled (micro) level, improvements were not uniform across tasks: for \EARLY{}, pooled \ac{AUROC} slightly decreased (0.75 vs.\ 0.77), whereas for \ABOVE{} it slightly increased (0.74 vs.\ 0.71). However, these differences are small and likely within estimation variability, so we refrain from over-interpreting the trends. When pooling treatment and control participants, \EARLY{} metrics were slightly lower than the corresponding logistic regression baselines (e.g., pooled \ac{AUROC} 0.75 vs.\ 0.77), while for \ABOVE{} both pooled \ac{AUROC} and \ac{AUPRC} increased (0.79 vs.\ 0.72 and 0.51 vs.\ 0.40, respectively). Figure~\ref{figC:plot_ROC} shows the \ac{ROC} curves for the \ac{CNN} models.

We additionally evaluated a preprocessing variant that normalizes each individual's signals per corresponding driving phase, to assess whether the models rely on phase-level distribution shifts across the study day (e.g., baseline drift or session-order effects) rather than impairment-related patterns. For both tasks, phase-wise normalization reduced performance compared to the standard preprocessing (e.g., \EARLY{} macro \ac{AUROC} \meanstd{0.82}{0.15} vs.\ \meanstd{0.88}{0.09}; \ABOVE{} macro \meanstd{0.81}{0.13} vs.\ \meanstd{0.86}{0.11}). This result suggests that phase-specific level shifts and longer-term temporal context carry an informative signal for intoxication detection, which is attenuated when each phase is normalized independently.

\begin{figure}[!htbp]
  \centering
  \includegraphics[width=0.8\linewidth]{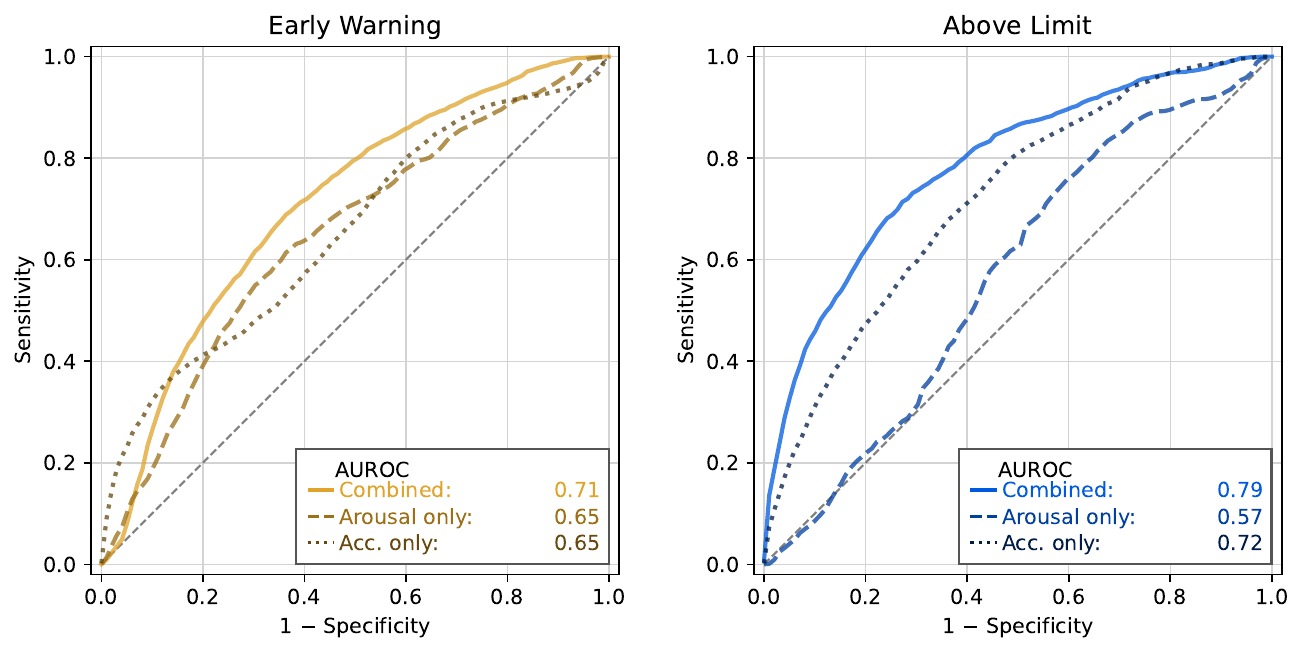}
    \caption{Receiver operating characteristic (ROC) curves for the \EARLY{} and \ABOVE{} tasks, comparing \ac{CNN} performance using physiological-arousal-only, accelerometer-only, and combined (physiological arousal and accelerometer) inputs. The dashed gray line indicates the performance of a random classifier.}
  \Description{TBD.} 
  ~\label{figC:plot_CNN_ablation}
\end{figure}

We further report the ablation results for the \ac{CNN}. Consistent with the logistic regression baseline, accelerometer-only models outperformed physiological-arousal-only models, particularly for \ABOVE{} (macro \ac{AUROC} \meanstd{0.84}{0.10} for accelerometer-only vs.\ \meanstd{0.61}{0.17} for physiological-arousal-only). However, the combined model achieved the best overall performance, indicating that physiological arousal estimates and wrist motion provide complementary information during driving. Figure~\ref{figC:plot_CNN_ablation} visualizes the results.

\begin{table*}[!htbp]
    \centering
    \Description{TBD} 
    \caption{Performance of the two-tower 1D \ac{CNN} for the \EARLY{} and \ABOVE{} binary alcohol-detection tasks under standard and per-phase normalization; as well as ablations of physiological-arousal-only and accelerometer-only (acc). We report macro-averaged (per-participant) and micro-averaged (pooled) \ac{AUROC} and \ac{AUPRC}, together with random \ac{AUPRC} baselines, for treatment participants and for the combined treatment+control sample.}
    \label{tabC:results_CNN}
    \begin{tabular}{cllcccc}
    \toprule
    & & & \multicolumn{2}{c}{\EARLY} & \multicolumn{2}{c}{\ABOVE} \\ 
    \toprule \noalign{\vskip 10pt}
    & & & standard & normalized & standard & normalized \\ [2pt]
    \cline{4-7}\noalign{\vskip 2pt}

    \multirow{3}{*}{\shortstack[c]{per-participant\\average}}
      & \multirow{3}{*}{treatment} & \ac{AUROC}
      & \meanstd{0.88}{0.09} & \meanstd{0.82}{0.15} & \meanstd{0.86}{0.11} & \meanstd{0.81}{0.13} \\
      & & \ac{AUPRC}
      & \meanstd{0.93}{0.05} & \meanstd{0.90}{0.09} & \meanstd{0.78}{0.17} & \meanstd{0.73}{0.16} \\
      & & \random{rand. \ac{AUPRC}}
      & \random{\meanstd{0.67}{0.04}} & \random{\meanstd{0.67}{0.04}} & \random{\meanstd{0.33}{0.04}} & \random{\meanstd{0.33}{0.04}} \\
    \hline

    \multirow{6}{*}{\shortstack[c]{pooled\\predictions}}
      & \multirow{3}{*}{treatment} & \ac{AUROC}
      & 0.75 & 0.69 & 0.74 & 0.68 \\
      & & \ac{AUPRC}
      & 0.85 & 0.82 & 0.63 & 0.53 \\
       & & \random{rand. \ac{AUPRC}}
      & \random{0.67} & \random{0.67} & \random{0.33} & \random{0.33} \\
    \cline{2-7}

      & \multirow{3}{*}{\shortstack[c]{treatment\\+ control}} & \ac{AUROC}
      & 0.71 & 0.65 & 0.79 & 0.65 \\
      & & \ac{AUPRC}
      & 0.54 & 0.54 & 0.51 & 0.30 \\
       & & \random{rand. \ac{AUPRC}}
      & \random{0.38} & \random{0.38} & \random{0.19} & \random{0.19} \\
    \hline \noalign{\vskip 15pt}
     & & & arousal only & acc only & arousal only & acc only \\[2pt]
     \cline{4-7}\noalign{\vskip 2pt}
    \multirow{3}{*}{\shortstack[c]{per-participant\\average}} & \multirow{3}{*}{treatment} & \ac{AUROC}
        & \meanstd{0.74}{0.17} & \meanstd{0.82}{0.14} & \meanstd{0.61}{0.17} & \meanstd{0.84}{0.10} \\
     & & \ac{AUPRC}
        & \meanstd{0.86}{0.11} & \meanstd{0.89}{0.09} & \meanstd{0.44}{0.20} & \meanstd{0.73}{0.16} \\
          & & \random{rand. \ac{AUPRC}}
          & \random{\meanstd{0.67}{0.04}} & \random{\meanstd{0.67}{0.04}} & \random{\meanstd{0.33}{0.04}} & \random{\meanstd{0.33}{0.04}} \\
    \hline
    \multirow{6}{*}{\shortstack[c]{pooled\\predictions}} & \multirow{3}{*}{treatment} & \ac{AUROC}
        & 0.71 & 0.64 & 0.60 & 0.70 \\
     & & \ac{AUPRC}
        & 0.82 & 0.77 & 0.38 & 0.53 \\
         & & \random{rand. \ac{AUPRC}}
          & \random{0.67} & \random{0.67} & \random{0.33} & \random{0.33} \\
    \cline{2-7}
     & \multirow{3}{*}{\shortstack[c]{treatment\\+ control}} & \ac{AUROC}
        & 0.65 & 0.65 & 0.57 & 0.72 \\
     & & \ac{AUPRC}
       & 0.50 & 0.58 & 0.21 & 0.39 \\
       & & \random{rand. \ac{AUPRC}}
          & \random{0.38} & \random{0.38} & \random{0.19} & \random{0.19} \\
    \hline
\end{tabular}
\end{table*}

\begin{figure}[!htbp]
  \centering
  \includegraphics[width=0.8\linewidth]{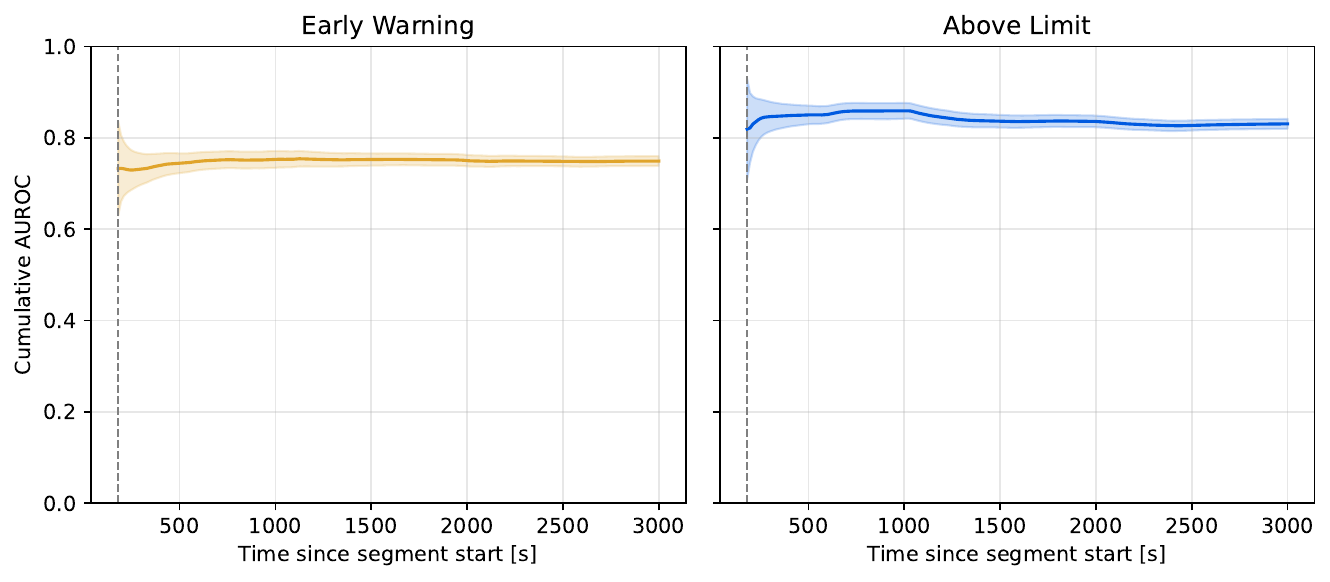}
    \caption{We temporally smooth model outputs by aggregating window-level predicted probabilities into 15\,s bins and computing a per-segment cumulative moving average (CMA) within each driving segment (i.e., each participant $\times$ phase). At each elapsed time $t$, we report the pooled \ac{AUROC} computed over all windows observed up to $t$. Shaded regions indicate 99\% DeLong confidence intervals. The dashed vertical line at 180\,s marks the earliest time at which a CMA-based prediction is available (one full window).}
  \Description{TBD.} 
  ~\label{figC:plot_time_smoothing}
\end{figure}

Furthermore, to approximate a deployment-relevant scenario, we evaluated temporally smoothed predictions by aggregating window-level predicted probabilities into 15 s bins and computing a per-driving-segment cumulative moving average. Figure~\ref{figC:plot_time_smoothing} reports pooled \ac{AUROC} as a function of elapsed time since segment start, computed cumulatively over windows observed up to time $t$. The resulting curves stabilize quickly, suggesting that reliable, continuously updated predictions can be obtained within the first $\approx$ 5 minutes of driving.

When reformulating the task as continuous \ac{BAC} regression rather than binary classification, the model achieved a mean absolute error of \gdL{0.019}. The Pearson correlation coefficient between predicted and reference \ac{BAC} values was 0.433, and the \ac{AUROC} obtained when using the regression outputs for the \ABOVE{} classification task was 0.746. Although this performance is lower than that of the dedicated binary \ac{CNN} classifier for \ABOVE{} detection (AUROC of 0.79), it remains within a comparable range. These results suggest that the model captures \ac{BAC}-associated signal variation and retains comparable discrimination for identifying driving above the \ac{WHO}-recommended limit.
At the same time, regression performance was likely constrained by the zero-inflated distribution of the training data.

\subsection{Categorical Phase Classification}

To further probe whether the learned representations capture alcohol-related impairment patterns rather than generic time-of-day or repeated-drive effects, we trained separate categorical \ac{CNN} models for the treatment and control groups to classify phases~1--3 (Table~\ref{tabC:results_multiclass}). For treatment participants, phase classification was substantially above random guessing (macro \ac{AUROC} \meanstd{0.87}{0.08}; macro \ac{AUPRC} \meanstd{0.78}{0.13} under standard preprocessing), whereas performance for control participants was markedly lower (macro \ac{AUROC} \meanstd{0.61}{0.13}; macro \ac{AUPRC} \meanstd{0.48}{0.12}) and close to the by-chance level of performance. These results are consistent with phase separability being driven primarily by alcohol-induced changes in the treatment group, with only limited phase-related (i.e., contemporal) patterns in the control group (e.g., due to fatigue, habituation, or other time-varying factors). When we applied per-phase normalization, it further reduced categorical classification performance in the control group, bringing it closer to random predictions, with a per-participant average AUROC of 0.55. In contrast, the treatment group performance remained strong, albeit slightly lower than the standard processing.

\begin{table*}[!htbp]
    \centering
    \Description{TBD} 
    \caption{Categorical phase-classification performance of the 1D \ac{CNN} within the treatment group ($n=31$) and the control group ($n=23$). We report macro-averaged (per-participant) and micro-averaged (pooled) \ac{AUROC} and \ac{AUPRC} under standard and per-phase normalization, together with random \ac{AUPRC} baselines.}
    \label{tabC:results_multiclass}
    \begin{tabular}{clcccc}
    \toprule
    & & \multicolumn{2}{c}{treatment ($n=31$)} & \multicolumn{2}{c}{control ($n=23$)} \\
     & & standard & normalized & standard & normalized \\
    \midrule
    \multirow{3}{*}{\shortstack[c]{per-participant\\average}} & \ac{AUROC}
        & \meanstd{0.87}{0.08} & \meanstd{0.82}{0.09} & \meanstd{0.61}{0.13} & \meanstd{0.55}{0.12} \\
     & \ac{AUPRC}
        & \meanstd{0.78}{0.13} & \meanstd{0.72}{0.12} & \meanstd{0.48}{0.12} & \meanstd{0.42}{0.09} \\
      & \random{rand. \ac{AUPRC}}
      & \random{\meanstd{0.33}{0.00}} & \random{\meanstd{0.33}{0.00}} & \random{\meanstd{0.33}{0.00}} & \random{\meanstd{0.33}{0.00}} \\
    \hline
    \multirow{3}{*}{\shortstack[c]{pooled\\predictions}} &  \ac{AUROC}
        & 0.80 & 0.78 & 0.56 & 0.54 \\
      & \ac{AUPRC}
        & 0.66 & 0.65 & 0.38 & 0.37 \\
       & \random{rand. \ac{AUPRC}}
      & \random{0.33} & \random{0.33} & \random{0.33} & \random{0.33} \\
    \hline
\end{tabular}
\end{table*}

\section{Discussion}
\label{secC:discussion}

\subsection{Summary of Findings}

Our results show that mobile, wearable-based drunk-driving detection based on wrist motion and physiological arousal estimation is feasible. In the following, we discuss several key aspects of the findings.

Detection of any alcohol level versus sober (\EARLY{}) works better than detection of whether participants are above or below the \ac{WHO}-recommended limit of \gdL{0.05}. This is consistent with previous work that reports the same pattern \cite{koch2023leveraging, Deuber2025}. The results indicate that human response to alcohol is more individual at higher \ac{BAC} levels (around \gdL{0.05}) than for the binary distinction sober vs.\ non-sober. From a technical perspective, class imbalance is also more pronounced for the \ABOVE{} task. When comparing the logistic regression approach with the \ac{CNN} approach, the picture is more nuanced. For \EARLY{} (the better-performing task), both model classes yield similar performance, suggesting that logistic regression already exploits most of the available separability. For \ABOVE{}, however, the \ac{CNN} improves on the logistic regression baseline. The window-length analysis further indicates that shorter windows may be sufficient for practical detection deployment, with only limited benefits from longer temporal context, especially compared to prior work \cite{koch2023leveraging}. This is consistent with Fig.~\ref{figC:plot_time_smoothing}, which suggests that temporally smoothed predictions reach stable discrimination after a short elapsed time. The additional regression analysis further suggests that the model captures meaningful continuous \ac{BAC}-related variation. However, because the present work focuses on classification, these results should primarily be viewed as complementary evidence and as indicating an avenue for future work on continuous intoxication estimation.

Across conditions, pooled predictions (micro-averages) consistently underperform macro-averages over participants. A plausible explanation is that macro \ac{AUROC}/\ac{AUPRC} only requires correct ranking within each participant (i.e., participant-specific score scales and operating points), whereas pooled evaluation additionally requires scores to be comparable across participants. In the latter, performance is determined by a single global ranking (and, for any fixed operating point, an implicit shared threshold) despite subject-specific differences. Consequently, these metrics are not directly comparable and highlight the need for improved cross-participant calibration. 

A design consideration is that the intoxication phases in the treatment group necessarily followed a fixed temporal order. Consequently, \ac{BAC} level, session index, accumulated driving experience, time-on-task, and circadian state are partially confounded by design. The placebo and reference groups were included to partially mitigate this limitation by matching the temporal structure without alcohol exposure, thereby reducing the risk that models rely on trivial differences between early and late sessions. As a further robustness analysis, we evaluated per-phase normalization, in which each participant's signals were z-normalized separately within each driving phase. If classification performance were primarily driven by coarse phase-level distribution shifts, performance would be expected to collapse under this normalization. Instead, per-phase normalization consistently reduces performance, albeit only slightly. This suggests that the models indeed exploit phase-level drifts over the study day, yet remain predictive even after such shifts are removed. To further probe this issue, we added a categorical phase-classification analysis in which we trained separate CNN models to classify phases 1--3 within the treatment and control groups. If generic repeated-drive, fatigue, or circadian effects dominated the learned representations, phase classification should also be strong among sober control participants. Instead, the findings further reinforce our model's ability to capture alcohol-related effects. The driving phase classification for control participants was slightly above chance, but separability between phases was significantly higher for the treatment group. This supports the interpretation that phase separability in the treatment group is not merely a generic artifact of session order.

Regarding modalities, the ablation analyses show that accelerometer features contribute more strongly than physiological arousal features, particularly for \ABOVE{}, while performance differences are smaller for \EARLY{}. This pattern is also consistent with the coefficient analysis of the logistic regression models, where mean absolute coefficients were generally larger for accelerometer-derived than for physiological arousal features, while differences between \EARLY{} and \ABOVE{} remained less pronounced. One interpretation is that higher levels of impairment are more strongly expressed in movement dynamics that are relevant for driving, which is consistent with reports that more severe intoxication is typically associated with greater impairment \cite{brumback2007effects}. This interpretation should be considered in light of unobserved variability in steering behavior. Participants were not constrained in their driving posture (e.g., one- vs. two-handed steering), and we did not record hand dominance or hand usage. While this preserves naturalistic driving behavior, steering style may influence wrist-worn accelerometer signals and contribute to inter-individual variability. Future work should therefore explicitly account for hand dominance, steering style, and device placement when interpreting motion-based features. At the same time, accelerometer and physiological arousal fusion yields the best overall performance, indicating that the two signals provide complementary evidence. A further interpretation is that we rely on a pretrained physiological arousal detection model and do not perform full end-to-end learning for physiological signals; this may slightly limit performance compared to accelerometer features. Conversely, the weaker physiological-arousal-only performance should not be interpreted as evidence that physiological sensing is uninformative in principle. Using a device-agnostic physiological arousal representation may improve portability across wearables that differ in access to raw \ac{PPG} versus only derived signals (e.g., \ac{IBI}). Future work could therefore evaluate end-to-end learning directly from raw \ac{PPG}, which may yield additional gains while retaining the option of physiological-arousal-based representations for cross-device generalization.

While direct comparisons are limited by differences in datasets, experimental protocols, and evaluation settings, our results are broadly in line with, and in some cases slightly higher than, reported performance in prior in-vehicle drunk-driving detection work using onboard sensors. The two most state-of-the-art studies in simulator and real-vehicle settings \cite{koch2023leveraging, Deuber2025} apply the same two binary detection tasks and report the same or somewhat lower performance using in-vehicle driver monitoring cameras and vehicle-based sensors. We achieved \meanstd{0.88}{0.09} and \meanstd{0.86}{0.11} for \EARLY{} and \ABOVE{}, respectively, while \NoHyper\citeauthor{koch2023leveraging}\endNoHyper\ reported \meanstd{0.88}{0.09} and \meanstd{0.79}{0.10} in a simulator setting \cite{koch2023leveraging}, and \NoHyper\citeauthor{Deuber2025}\endNoHyper\ achieved \meanstd{0.84}{0.11} and \meanstd{0.80}{0.10} in a real vehicle using in-vehicle sensing \cite{Deuber2025}. Although direct comparison is limited (e.g., due to lack of \ac{LOSO} evaluation, different task formulations, or other methodological differences), our results are broadly comparable to mobile drunk-driving detection outside of driving contexts and highlight wearable-based sensing as a viable alternative to in-vehicle approaches, with the advantage of scaling to drivers in older vehicle fleets without dedicated in-cabin sensing hardware.

\subsection{Practical Relevance}

Wearable-based intoxication detection is practically relevant because it directly targets a key failure mode: many drivers miscalibrate their own impairment and underestimate their objective intoxication \cite{love2025systematic, kochling2021hazardous,laude2016drivers, amlung2014effects}. The primary value of a smartwatch-based detector is not to replace legal enforcement or breath testing, but to reduce this awareness gap by providing objective feedback in situations where subjective judgment is unreliable. Because the proposed approach relies only on a widely available wrist-worn device, it is inherently scalable and can be deployed without additional vehicle hardware, proprietary telemetry access, or in-cabin cameras. This makes it a pragmatic alternative to vehicle-centric systems: it can reach drivers in older vehicle fleets and across heterogeneous mobility contexts, while remaining non-invasive and comparatively low-cost. A second practical implication follows from driver-state-aware assistance systems. If impairment sensing is available, safety functions can, in principle, adapt their behavior to the driver’s condition. In this context, a wearable-derived signal can act as an auxiliary indicator that the driver’s state has changed in a manner consistent with alcohol exposure, and thus that impairment is likely.

The results also suggest a deployment-relevant design direction: while mean performance improves only slightly with longer temporal aggregation, the performance confidence interval narrows, indicating more stable predictions, with reliable, continuously updated estimates available within $\approx$5 minutes. Because nuisance alarms would be unacceptable, the most realistic operationalization is a temporally smoothed risk score that integrates evidence over longer durations rather than triggering on single windows. Such a score could support the two complementary intervention pathways discussed above: (1) user-facing feedback and (2) vehicle-side safety system adaptation. Both pathways emphasize prevention and harm reduction rather than adjudicating legal impairment.

\subsection{Contributions}

This work makes three core contributions. First, it demonstrates the feasibility of detecting alcohol-impaired driving using only consumer-grade smartwatch sensing in a realistic driving context, without relying on specialized transdermal sensors or proprietary in-vehicle integration. This lowers cost and access barriers and positions wearable-based sensing as a pragmatic complement to vehicle-centric approaches. Importantly, our goal is risk-aware, symptom-based impairment detection: we identify wearable-derived signals associated with alcohol-impaired driving that can support preventive feedback and driver-state-aware assistance systems, rather than attempting causal inference or diagnostic attribution of intoxication. Second, it shows the development and evaluation of an off-the-shelf smartwatch-based drunk-driving detection system in a real vehicle on a closed test track. The study uses a three-arm study design (treatment, placebo, open-label reference) not as a methodological contribution in itself, but to strengthen the interpretation of wearable-derived impairment signals by accounting for expectancy, fatigue, and time-on-task confounds. Third, we show generalization to unseen drivers by consistently applying \ac{LOSO} cross-validation. In addition, we provide a structured evaluation across model families and experimental factors, including a state-of-the-art baseline and a two-tower 1D \ac{CNN}, alongside analyses of window size, modality contributions (accelerometry versus physiological arousal), normalization, and categorical phase classification to probe alcohol-specific structure. Taken together, these contributions help close the gap between prior wearable intoxication studies that largely occur outside the driving context and prior drunk-driving detection work, which relies primarily on in-vehicle sensors, and they establish a foundation for future on-road validation and deployment-oriented research.




\subsection{Limitations and Outlook}

While our findings are promising, several limitations could constrain generalization and highlight opportunities for future work. First, while the sample size is substantial for a controlled alcohol administration study, it does not capture the full diversity of real-world drivers in terms of demographics, health status, driving styles, and wearable usage habits. Larger, more heterogeneous studies are needed to assess generalizability to the broader driving population. Second, the current labels reflect alcohol exposure, but the specificity of the learned signal is not yet established: other driver states (e.g., fatigue, distraction, non-alcohol-related physiological arousal, illness, or medication effects) may induce partially similar physiological or movement changes and could give rise to false positives.

Third, for legal and ethical reasons, this study deliberately uses a closed-track design, which represents the highest-fidelity setting currently permissible for controlled alcohol exposure. Although the protocol was designed to approximate real driving, open-road conditions would likely introduce additional variability (e.g., route heterogeneity) and a corresponding distribution shift.  Thus, while the test-track design provides high internal validity, its ecological validity remains limited. Driving took place during daytime, without surrounding traffic, secondary non-driving tasks, or externally induced distractions. In addition, the presence of a licensed driving instructor with dual pedals may have created a supervised driving context, which could both increase rule-compliant driving and reduce perceived accident risk. Finally, all data were collected in the same vehicle model, which limits conclusions about generalization across vehicle types, seating positions, steering dynamics, and cabin layouts. Accordingly, future work should prioritize on-road validation under naturalistic conditions, for example, via observational studies generating a sober validation dataset. Such data would make it possible to quantify distribution shift, assess false-positive rates in everyday driving, and evaluate whether calibration or personalization strategies are needed before deployment. These studies would increase ecological validity, although at the cost of reduced experimental control and the absence of controlled intoxication labels. In the longer term, wearable sensing could also be integrated into large-scale naturalistic driving studies, where extensive real-world driving data is collected and rare naturally occurring impaired-driving episodes may be observed under appropriate ethical and legal safeguards.

Moreover, while the \ac{CNN} improves per-participant discrimination, the gap between macro and micro performance indicates that cross-participant calibration is a deployment-relevant consideration. Macro-averaged metrics primarily reflect whether the model can rank intoxicated and sober windows within the same participant, whereas pooled metrics more closely approximate a cold-start deployment scenario in which a common model and threshold are applied to previously unseen drivers. This distinction is important because wearable sensing models often suffer from inter-individual variability in physiology, movement patterns, device placement, and baseline signal distributions. Similar challenges have been discussed in prior mobile and wearable sensing work, where personalization and generalization are treated as key requirements for robust real-world deployment~\cite{banovic2018warming,grammenos2018sensing,sztyler2017online,meegahapola2022generalization}. For practical smartwatch-based drunk-driving detection, this suggests that a purely population-level model may be most appropriate as an initial cold-start model, but that performance could likely benefit from personalization. Promising strategies include collecting a short sober driving baseline, normalizing risk scores relative to typical physiological and motion profile of an individual, adapting decision thresholds per user, or incrementally updating calibration parameters as more user-specific data become available. Future work should therefore evaluate explicit calibration and personalization strategies and quantify how much user-specific data is needed to close the macro--micro performance gap.

More broadly, future studies could further build on this foundation by exploring additional time-series \ac{ML} approaches under the same \ac{LOSO} scheme and assess whether they yield incremental robustness gains. Likewise, temporal postprocessing (e.g., rolling aggregation, hysteresis thresholds, Bayesian filters, or sequence models) may provide a pragmatic way to reduce jitter and improve decision stability in real time. Another promising direction is to leverage wearable foundation models (e.g., \cite{narayanswamy2024scaling}) pretrained on large-scale unlabeled or weakly labeled data (potentially lower quality) and then fine-tuned on high-quality experimental datasets.

A practical limitation of the present approach concerns detection latency. Our current pipeline uses 180 second windows, and when combined with temporal smoothing, stable predictions emerge after approximately the first 5 minutes of driving. The 180 second window length was selected a priori based on prior literature rather than optimized on the present data, to reduce the risk of overfitting. At the same time, our a posteriori analysis across multiple window lengths suggested that longer windows yielded only limited performance gains. While such window lengths may be acceptable for longer trips, they may reduce applicability in short-distance driving scenarios. Future work should therefore investigate whether shorter windows can retain sufficient predictive performance while enabling earlier detection.

Moreover, a limitation is the domain shift introduced by applying the physiological arousal model of \citet{mishra2020evaluating} outside the context in which it was developed. The model was trained and evaluated on controlled physiological arousal paradigms, including mental, startle-based, and physical stressors, whereas our setting combines alcohol exposure, active driving, task demands, fatigue, and safety supervision. We therefore interpret this signal as a proxy for autonomic arousal rather than an alcohol-specific biomarker. Our experimental design was intended to partially address this specificity challenge. The placebo and reference groups completed the same three-session schedule while remaining sober, which helps separate alcohol-related effects from generic time-on-task, repeated-driving, and circadian effects. In addition, the categorical phase-classification analysis and the per-phase normalization analysis suggest that phase separability was stronger under alcohol administration than under the sober control schedules. Nevertheless, these analyses cannot rule out all non-alcohol sources of physiological arousal. Future work should therefore investigate end-to-end models trained directly on raw or minimally processed physiological signals, rather than relying on an intermediate arousal-estimation model developed in a different context.


A further direction for future work is multimodal fusion between wearable and in-vehicle sensing. The present paper focuses on smartwatch-only detection, as this setting is the most scalable and does not depend on proprietary vehicle integration or additional onboard hardware. Nevertheless, combining wearable signals with vehicle-based measurements may further improve detection performance and robustness, particularly if the two modality groups capture complementary aspects of alcohol-related impairment. Future work should therefore compare smartwatch-only, vehicle-only, and fused models to assess the added value of multimodal integration.

Finally, privacy and ethical considerations are central for wearable-based impairment sensing, as continuous physiological monitoring is inherently sensitive. Future work should therefore reinforce data-minimization strategies, prioritize on-device processing, ensure transparency to users, and define clear governance around when and how risk scores are shared with vehicles or third parties, alongside safeguards to prevent misuse.

\section{Conclusion}
\label{secC:conclusion}

Alcohol-impaired driving remains a major, yet preventable, cause of road traffic harm, and many drivers misjudge their own intoxication, motivating accessible, non-invasive tools that close this awareness gap without requiring additional in-vehicle hardware. In this work, we use consumer-grade smartwatches in a randomized, controlled test-track study to collect wrist motion and physiological arousal signals across sober and intoxicated driving phases, and train both LASSO-regularized logistic regression and two-tower 1D \ac{CNN} models under \ac{LOSO} validation for two binary detection tasks (\EARLY{} and \ABOVE{}). Our results show that smartwatch-based sensing can reliably distinguish sober from alcohol-impaired driving, with performance clearly above chance and broadly comparable to in-vehicle approaches, thereby supporting the premise that the motivation articulated in the introduction can be addressed using widely available wearable devices. Overall, this work establishes wearable-based drunk-driving detection as a viable and scalable complement to vehicle-centric systems and legal enforcement, while highlighting the need for further research on calibration, real-world deployment, and ethical safeguards.

\bibliographystyle{ACM-Reference-Format}
\bibliography{bib}




\end{document}